%% file: main.tex
\documentclass{article}
\input{packages}
\input{biblio}

\title{Data-driven prediction of tool wear using Bayesian regularized artificial neural networks}

\input{authors}
\date{}

\begin{document}
\nolinenumbers
\maketitle
\input{sections/00.abstract}
\input{sections/01.introduction}

\section{Data collection}

In order to estimate the tool wear using the BRANN model, four different data sets have been collected. The data available in each data set are produced by performing experiments. Details about each data set are given in the following subsections.

\subsection{NASA Ames milling data set}
\begin{figure}
	\centering
	\includegraphics[scale=0.8]{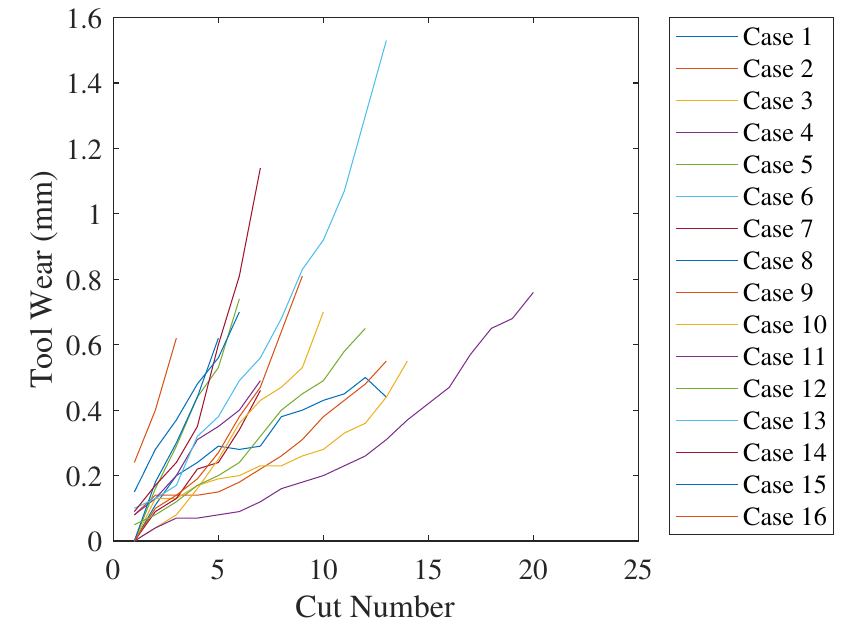}
	\caption{Tool wear for all sixteen cases in the NASA Ames milling data set.}
	\label{Fig.1}
\end{figure}

The data were produced by performing milling of stainless steel (J45) and cast iron using  70 mm face mill cutter having 6 inserts of tungsten carbide (WC) coated with TiC/TiC-N/TiN in sequence. The experiments were performed at a cutting speed of 200 m/min, depth of cuts of 0.75 and 1.5 mm, and feeds of 0.25 mm/rev and 0.5 mm/rev. A total of 16 sets of experiments were carried out at different combinations of process parameters with varying number of cuts, as shown in Table \ref{tab1}. A vibration sensor and acoustic emission sensor were mounted to monitor the signals. The signals are then processed and used to monitor the tool wear. The flank wear (Vb) was observed and measured using a microscope for each cutting condition. Fig. \ref{Fig.1} shows the results for Vb for all 16 cutting conditions. The data generated shows that the number of cuts was not constant for each cutting condition. It also seems that the flank wear was not always measured and at times when no measurements were taken. Detailed information regarding the dataset can be found in \cite{Agogino2007} or at https://www.nasa.gov/content/prognostics-center-of-excellence-data-set-repository.
\begin{table}[htbp]
	\centering
 
	\caption{Experimental conditions of the NASA Ames milling data set.}
	\begin{tabular}{ccclc}
		\hline
		Case  & Depth of cut & Feed  & Material \\
		\hline
		1     & 1.5   & 0.5   & 1 – cast iron  \\
		\hline
		2     & 0.75  & 0.5   & 1 – case iron \\
		\hline
		3     & 0.75  & 0.25  & 1 – cast iron \\
		\hline
		4     & 1.5   & 0.25  & 1 – cast iron \\
		\hline
		5     & 1.5   & 0.5   & 2 – steel \\
		\hline
		6     & 1.5   & 0.25  & 2 – steel \\
		\hline
		7     & 0.75  & 0.25  & 2 – steel \\
		\hline
		8     & 0.75  & 0.5   & 2 – steel \\
		\hline
		9     & 1.5   & 0.5   & 1 – cast iron \\
		\hline
		10    & 1.5   & 0.25  & 1 – cast iron \\
		\hline
		11 & 0.75 & 0.25 & 1 – cast iron  \\
		\hline
		12 & 0.75 & 0.5 & 1 – cast iron  \\
		\hline
		13    & 0.75  & 0.25  & 2 – steel  \\
		\hline
		14    & 0.75  & 0.5   & 2 – steel  \\
		\hline
		15 & 1.5 & 0.25 & 2 – steel  \\
		\hline
		16 & 1.5 & 0.5 & 2 – steel  \\
		\hline
	\end{tabular}%
	\label{tab1}%
\end{table}%

\subsection{2010 PHM Data Challenge data set}
 The data were produced by performing the milling of stainless steel (HRC 52) using a WC cutter having a three-flute ball nose. The spindle speed was 10400 rpm, the feed was 1555 mm/min, axial and radial depth of cut were 0.2 mm and 0.125 mm, respectively. A Kistler 3-component dynamometer was mounted to measure the machining forces and an accelerometer was mounted to measure the machine tool vibrations. In addition, an acoustic emission sensor was employed to monitor the high-frequency stress wave generated during the machining process. An optical microscope was used to observe the flank wear on each flute of the cutter. For further details about the dataset, readers are recommended to refer to https://www.phmsociety.org/competition/phm/10.
\subsection{Dataset of NUAA Ideahouse}
The data set was produced to analyze the influence of process parameters on the tool wear by monitoring the signals in the milling operation \cite{li2021tool}. The experiments were performed on titanium alloy (TC4) and superalloy using different tool materials and at varying process parameters. The details of the experiments are presented in Table \ref{tab2}. Cutting force, vibration, spindle current, and power were monitored to examine the tool wear during the milling process. The monitoring signals were collected by sensors and the tool wear was observed by scanning electron microscope (SEM) after milling of each pocket layer. The maximum width of the flank wear land was chosen as per the ISO standards for tool wear criteria. In this dataset, the Vb is estimated at different spindle speeds, axial and radial depth of cuts, and workpiece materials.

\begin{figure}
	\centering
	\includegraphics[scale=1]{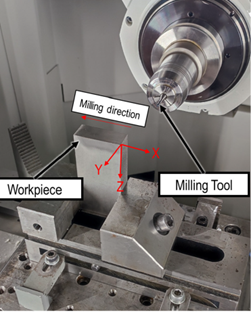}
	\caption{Experimetal setup for end milling \cite{hojati2022prediction}.} 
	\label{Figex}
\end{figure}

\begin{landscape}

\begin{table}[htbp]
	\centering
	\caption{The details of the cutting experiments of the NUAA Ideahouse dataset.}
	\begin{tabular}{cccccp{22.07em}cc}
		\hline
		No.   & fz (mm/r) & n (r/min) & ap (mm) & Tool diameter (mm) & \multicolumn{1}{c}{Tool material} & Workpecice material   \\
		\hline
		1     & 0.045 & 1750  & 2.5   & \multirow{9}[18]{*}{12} & \multirow{9}[18]{*}{Carbide Endmill without rounded corner} & \multirow{9}[18]{*}{Titanium alloy} \\
		\cline{1-4}    2     & 0.045 & 1800  & 3     &       & \multicolumn{1}{c}{} &       &  \\
		\cline{1-4}    3     & 0.045 & 1850  & 3.5   &       & \multicolumn{1}{c}{} &       &  \\
		\cline{1-4}    4     & 0.05  & 1750  & 3     &       & \multicolumn{1}{c}{} &       &  \\
		\cline{1-4}    5     & 0.05  & 1800  & 3.5   &       & \multicolumn{1}{c}{} &       &  \\
		\cline{1-4}    6     & 0.05  & 1850  & 2.5   &       & \multicolumn{1}{c}{} &       &  \\
		\cline{1-4}    7     & 0.055 & 1750  & 3.5   &       & \multicolumn{1}{c}{} &       &  \\
		\cline{1-4}    8     & 0.055 & 1800  & 2.5   &       & \multicolumn{1}{c}{} &       &  \\
		\cline{1-4}    9     & 0.055 & 1850  & 3     &       & \multicolumn{1}{c}{} &       &  \\
		\hline
		10    & 0.06  & 2200  & 6     & 12    & \multirow{2}[4]{*}{Carbide Endmill without rounded corner} & \multirow{3}[6]{*}{Titanium alloy} \\
		\cline{1-5}    11    & 0.07  & 1300  & 8     & 20    & \multicolumn{1}{c}{} &       &  \\
		\cline{1-6}    12    & 0.08  & 500   & 3     & 12    & High speed steel &       &  \\
		\cline{1-7}    13    & 0.07  & 1200  & 4     & 12    & Carbide Endmill without rounded corner & Superalloy &  \\
		\hline
	\end{tabular}%
	\label{tab2}%
\end{table}%

\subsection{Dataset of inhouse performed end milling of Ti6Al4V}

In this investigation, the end milling experiments were performed on Ti6Al4V taking TiAlN coated WC end mill cutter (Make: HB microtec GmBH, KG, Germany) having diameter of 3 mm and 3 flutes. This study used a five axis CNC machine tool (Haas-Multigrind CA, Trossingen, Germany) with a Siemens Sinumeric controller. The milling tests carried out at cutting speeds of 50-95 m/min, feed per tooth of 17-50 µm/tooth, and axial and radial depth of cut was 1 mm and 3 mm respectively. A total of 15 experiments were conducted under dry conditions. The dynamometer was employed to collect the force data from $x$, $y$ and $z$ directions, as shown Fig. \ref{Figex}. It can be seen that the tool moved from the right to left side of the workpiece. Since the slots to be created with a depth of 6 mm, six milling passes each with axial depth of cut of 1 mm were performed. For each combination of process parameters, six passes were performed with tool having no significant tool wear. For some combinations, the tests were stopped when the tool was found broken, as shown in Table \ref{tab15}. It can be observed that total 4 combinations show the tool breakage out of 15. Furthermore, the tool wear was predicted based on the force data. In order to classify the tool condition, the tool wear criteria given in ISO 3685:1993 was considered, that is, the average tool wear should not be excided 0.3 mm. The tool conditions was predicted and compared with experimental conditions.

\end{landscape}

\section{Method}

\subsection{Artificial Neural Networks}

Artificial Neural Network (ANN) is known as a computational model inspired by the structure and function of the biological neural networks in the human brain \cite{agatonovic2000basic}. It is a type of machine learning (ML) algorithm that is capable of learning and performing tasks such as pattern recognition, classification, regression, and more \cite{abiodun2018state}.

\begin{figure*}[!htb]			
	\centering		
	\begin{subfigure}{0.45\textwidth}
		\centering
		\includegraphics[scale = 0.3]{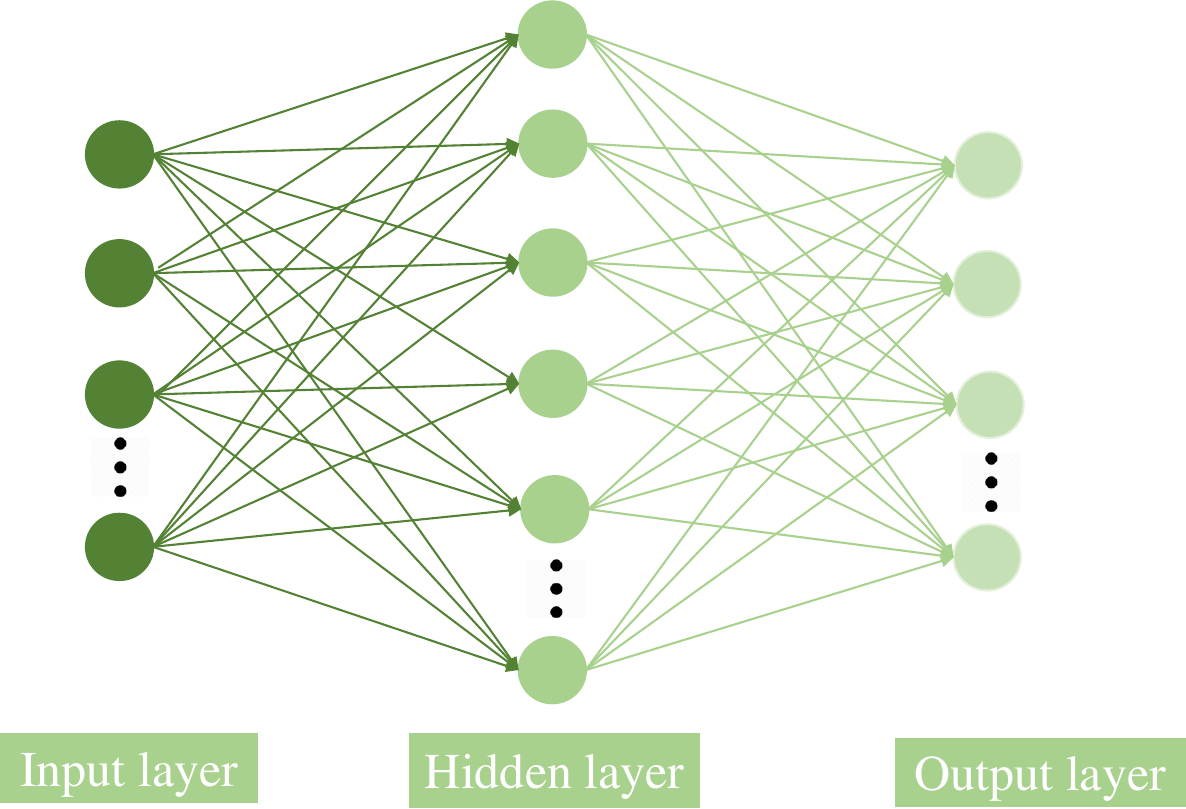}
		\caption{A typical architecture of an ANN model.}
		\label{Fig.2a}	
	\end{subfigure}
	~
	\begin{subfigure}{0.45\textwidth}
		\centering
		\includegraphics[scale = 0.3]{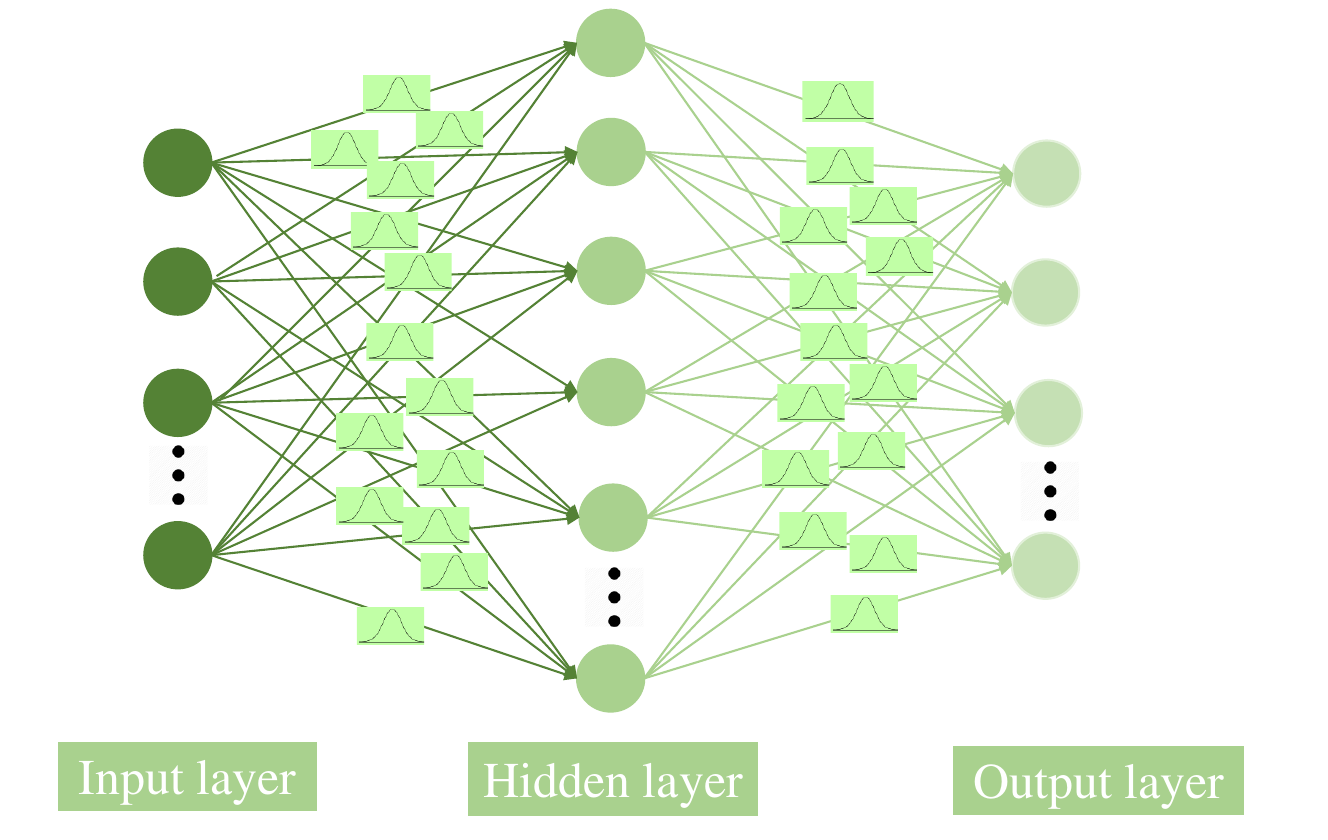}
		\caption{A typical architecture of the BRANN model.}
		\label{Fig.2b}	
	\end{subfigure}
	\caption{ A typical architecture of neural networks.}		
	\label{Fig.11}		
\end{figure*}

The basic building block of an artificial neural network is an artificial neuron, also known as a perceptron. It takes multiple input signals, applies weights to these inputs, computes a weighted sum, and passes the result through an activation (transfer) function to produce an output. The activation function introduces non-linearity and allows the network to model complex relationships between inputs and outputs. Multiple artificial neurons are interconnected to form layers within an ANN. Typically, an ANN consists of an input layer, one or more hidden layers, and an output layer. The neurons in each layer are connected to the neurons in the subsequent layer, forming a network of interconnected nodes. A typical architecture of an ANN model is depicted in Fig. \ref{Fig.2a}. The computing process of an ANN can be expressed by a mathematical formulation as follows \cite{hassoun1995fundamentals}:
\begin{equation}
    {\bf{y}} = f({\bf{Wx}} + b)
\end{equation}
where {\bf{y}} is the output vector; $f$ is the activation function; {\bf{x}} is the input vector; {\bf{W}} and $b$ are the weight matrix and bias parameter, respectively.

During the training phase of an ANN, the network adjusts the weights of the connections between neurons based on a specified loss (objective) function. In the context of regression problems, it is common to employ the mean square error (MSE) function as a loss function for the ANN model. The mathematical expression for MSE is typically defined as \cite{karunasingha2022root}:

\begin{equation}
    L_{MSE} = \frac{1}{N}\sum_{i=1}^{N}\left(Y_i^T-Y_i^P\right)^2
\end{equation}
where $Y_i^T$ represents the target corresponding to sample $i$, while $Y_i^P$ refers to the output of the model for sample $i$. The variable $N$ denotes the total number of samples.

The training process of an ANN is implemented by a backpropagation algorithm, which computes the gradient of the loss function with respect to the network’s weights and updates them accordingly. The objective is to minimize the difference between the predicted output of the network and the desired output (i.e., minimize the MSE loss function). Once the best-trained model is obtained, an accurate output corresponding to a given input can be quickly predicted from the trained model.

\subsection{Bayesian Regularized Artificial Neural Networks}
Being different from the traditional ANN, Bayesian regularized ANN (BRANN) combines the principles of Bayesian inference with ANN \cite{mackay1992bayesian}. BRANN introduces Bayesian regularization into the training process by adding an additional term into the loss function. This additional term penalizes the presence of large weights, which is introduced to provide a smoother network response. The loss function introduced in BRANN is rewritten as follows \cite{foresee1997gauss}:

\begin{equation}
    L = \beta\frac{1}{N}\sum_{i}^{N}\left(Y_i^T-Y_i^P\right)^2+\alpha\frac{1}{N}\sum_{i}^{N}w_i^2
\end{equation}
where $w_i$ is the network's weight; $\alpha$ and $\beta$ are the hyperparameters of the loss function. If $\alpha\ll\beta$, the training algorithm will prioritize reducing errors, resulting in smaller error values. Conversely, if $\alpha\gg\beta$, training will prioritize reducing the size of weights, even if it comes at the cost of increased network errors.

In the BRANN, the weights of the neural network are considered random variables instead of fixed values. After the data is retrieved, the density function for the ANN weights can be updated according to Bayes' rule as follows \cite{foresee1997gauss}:

\begin{equation}
    P\left(w\middle|D,\alpha,\beta,M\right)=\frac{P\left(D\middle|w,\beta,M\right)P\left(w\middle|\alpha,M\right)}{P\left(D\middle|\alpha,\beta,M\right)}
\end{equation}
in which $D$ is the training data set; $M$ is a particular ANN model used;  $P\left(D\middle|w,\beta,M\right)$ denotes the likelihood function which quantifies the probability of the observed occurrence given the network weights; $P\left(w\middle|\alpha,M\right)$ defines the prior density which characterizes our knowledge or beliefs about the weights before any data is collected; The term $P(D|\alpha, \beta, M)$ serves as a normalizing factor to ensure that the total probability sums up to 1. An illustration of the computing process of a typical BRANN model is depicted in Fig. \ref{Fig.2b}.

By using Bayesian inference, BRANN can estimate the posterior distribution over the network's parameters rather than finding a single-point estimate. In this Bayesian framework, the optimal weights are determined by maximizing the posterior probability $P(w|D, \alpha, \beta, M)$. This maximization process is equivalent to minimizing the regularized loss function $L$.

The Bayesian regularization in BRANN helps to prevent overfitting by imposing a penalty on large weights, encouraging a more robust and generalized model. Additionally, the uncertainty estimates provided by BRANN can be useful in decision-making processes that require quantifying the uncertainty in predictions. BRANN has been applied in various domains, including regression, classification, and reinforcement learning tasks \cite{bharadiya2023review}. In this study, the BRANN is applied to predicting the milling tool wear.

\subsection{Application of Bayesian Regularized Artificial Neural Networks for tool wear prediction}
\begin{figure}
	\centering
	\includegraphics[scale=0.5]{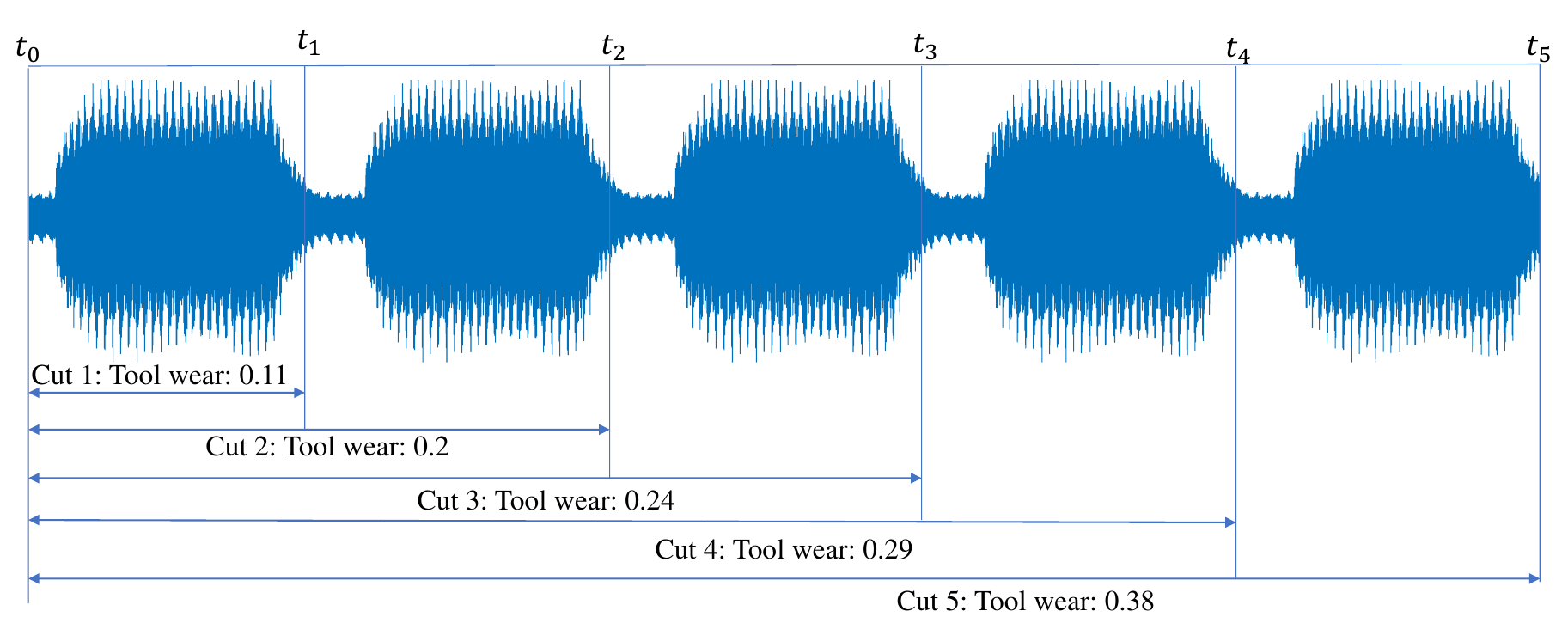}
	\caption{Determining of a particular input signal for each cut corresponding to the tool wear.}
	\label{FigProceesSignal}
\end{figure}

In order to predict the tool wear using the BRANN, both process parameters (e.g., feed rate, deep of cut) and monitoring sensor signals (e.g., cutting force, torque, vibration, acoustic emission (AE), spindle power, current) are considered as inputs of the BRANN model. Since the long signals are collected from sensors for each cutting condition, a feature extraction process is performed before embedding these signals into the model with the aim of reducing the computational cost in the training process. In addition, the tool wear is measured after each cut and the tool wear at the current cut is accumulated from the preceding cut. Therefore, in order to predict the tool wear at cut $n$, we use the signal measured from time $t_{0}$ to time $t_{n}$. A detailed description of how the input signal for each cut is determined can be found in Fig. \ref{FigProceesSignal}. Since the length of input signals for each cut varies, three statistical features, including the minimum, maximum, and mean of each input signal are extracted. These extracted features are then incorporated with the process parameters to form the final input feature vector of the model. Furthermore, in order to ensure all the inputs have an equal contribution during the model's training process, the original input features are normalized by the MinMaxscaler. The mathematical expression of the MinMaxscaler is presented as follows \cite{kramer2016scikit}:

\begin{equation}
    x^{norm}=\frac{x-x^{min}}{x^{max}\ -x^{min}}
\end{equation}
where $x^{norm}$ defines the normalized variable; $x$ is the original variable; $x^{min}$ and $x^{max}$ are minimum and maximum variables, respectively.

After the data preprocessing and normalization, all the input features are fed into the BRANN model for training. The training process is implemented through backpropagation learning. The model is trained until the stopping criterion is reached. In this study, the model is trained until the maximum training epoch reaches 1000 or the criterion of early stopping is reached. 

Moreover, in order to evaluate the performance of the BRANN in predicting the tool wear, three different well-known metrics including mean absolute error (MAE), root mean square error (RMSE), and the coefficient of determination ($R^2$) are used here. The mathematical formulation of the MAE, RMSE, and $R^2$ are expressed as follows:

\begin{equation}
    MAE\ =\frac{1}{N}\sum_{i=1}^{N}\left|Y_i^T-Y_i^P\right|
\end{equation}

\begin{equation}
    RMSE\ =\ \sqrt{\frac{1}{N}\sum_{i=1}^{N}\left(Y_i^T-Y_i^P\right)^2}
\end{equation}

\begin{equation}
    R^2=1-\frac{\sum_{i=1}^{N}\left(Y_i^T-Y_i^P\right)^2}{\sum_{i=1}^{N}\left(Y_i^T-{\bar{Y}}_i\right)^2}
\end{equation}
in which ${\bar{Y}}_i$ is the mean value of the targets.

\section{Results and Discussions}

In this section, the performance and applicability of the proposed BRANN model in predicting tool wear are examined by considering four distinct experimental data sets. Initially, the NASA Ames milling dataset including various operating conditions is considered. Subsequently, the proposed model is verified using the 2010 PHM Data Challenge dataset. Next, the proposed BRANN is validated by the NUAA Ideahouse tool wear dataset. Lastly, an inhouse performed end milling of Ti6Al4V dataset is taken into account to further validate the geralization of the proposed model. To demonstrate the accuracy and reliability of the proposed model, the results achieved by the proposed model are compared with those obtained from other existing models such as LR, SVR, MLP, CNN, LSTM, and MIFS. Furthermore, the impact of individual input features, training data size, hidden units, training algorithms, and transfer functions on the performance of the proposed model is also investigated.

\subsection{Prediction of tool flank wear based on NASA Ames milling data set}
\label{section 4.1}
In this case, the performance and applicability of the BRANN are validated using the NASA Ames milling data set. The process information and monitoring signals serve as input parameters for the model, while the output of the model is the flank wear of the tool. A comprehensive description of the input and output information of the model is found in Table \ref{tab3}.

\begin{table}[]
\centering
\caption{The information of inputs and output of the BRANN model for predicting the tool wear using the NASA Ames milling data set.}
\begin{tabular}{cclll}
\hline
\multicolumn{1}{l}{}       & No. &  & Measured parameters & Description                     \\ \hline
\multirow{8}{*}{Inputs}    & 1   &  & DOC                 & Depth of cut                  \\ \cline{2-5} 
                           & 2   &  & FEED                & Feed rate                       \\ \cline{2-5} 
                           & 3   &  & SMCAC               & AC spindle motor   current      \\ \cline{2-5} 
                           & 4   &  & SMCDC               & DC spindle motor   current      \\ \cline{2-5} 
                           & 5   &  & TableVibration      & Table vibration                 \\ \cline{2-5} 
                           & 6   &  & SpindleVibration    & Spindle   vibration             \\ \cline{2-5} 
                           & 7   &  & AeAtTable           & Acoustic   emission at table    \\ \cline{2-5} 
                           & 8   &  & AeAtSpindle         & Acoustic   emission at spindle  \\ \hline
\multicolumn{1}{l}{Ouputs} & 1   &  & Flank wear (VB)     & Flank wear, measured after runs \\ \hline
\end{tabular}
\label{tab3}%
\end{table}

\begin{figure}
	\centering
	\includegraphics[scale=0.6]{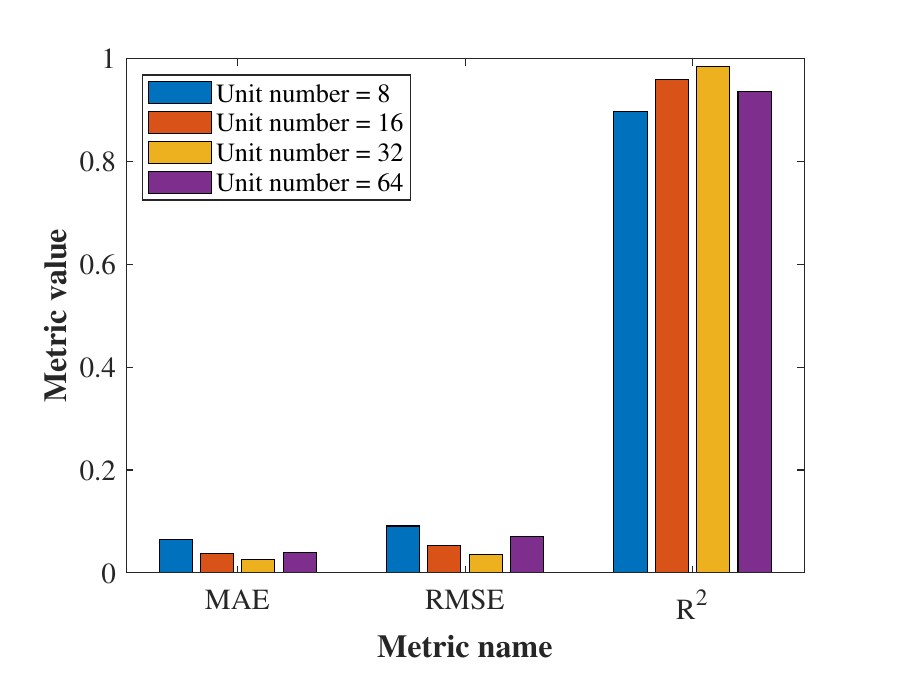}
	\caption{Comparison of the performance of the proposed BRANN model with different numbers of hidden units using the NASA Ames milling data set.}
	\label{Fig.3}
\end{figure}

Firstly, various investigations have been conducted to determine an appropriate number of hidden units in the BRANN, as depicted in Fig. \ref{Fig.3}. The results reveal that the BRANN model with 32 hidden units exhibits the lowest values of MAE and RMSE, along with the highest value of $R^2$, outperforming other configurations of hidden units. This implies that the BRANN with 32 hidden units achieves the highest accuracy in predicting tool wear compared to the other investigated options. Consequently, the proposed BRANN model is constructed with 32 units in the hidden layer for tool wear prediction using the NASA Ames milling data set.

Additionally, the selection of an appropriate transfer function for the network is crucial. Table \ref{tab4} presents the investigation results of the proposed BRANN model with different transfer functions for predicting tool wear using the NASA Ames milling data set. It is evident that the tansig and elliotsig transfer functions yield exceptional results in predicting tool wear compared to the compet, hardlim, logsig, poslin, purelin, radbas, satlin, and tribas functions. Moreover, although both tansig and elliotsig functions perform well in predicting tool wear, the tansig function demonstrates superior performance in the test set compared to the elliotsig function. Therefore, the tansig function is selected as the transfer function in the proposed BRANN model.

\begin{table}[]
 \centering
	\caption{Comparison of the performance of the proposed BRANN model with different transfer functions for predicting tool wear using the NASA Ames milling data set.}
\begin{tabular}{lccccccl}
\hline
\multirow{2}{*}{Transfer functions} & \multicolumn{2}{c}{MAE}           & \multicolumn{2}{c}{RMSE}          & \multicolumn{2}{c}{$R^2$}             &  \\ \cline{2-8} 
                                    & Training        & Test            & Training        & Test            & Training        & Test            &  \\ \hline
\textbf{tansig}                     & \textbf{0.0040} & \textbf{0.0268} & \textbf{0.0060} & \textbf{0.0365} & \textbf{0.9994} & \textbf{0.9842} &  \\ \hline
compet                              & 0.1788          & 0.1803          & 0.2352          & 0.2317          & 0.1702          & 0.0207          &  \\ \hline
elliotsig                           & 0.0031          & 0.0277          & 0.0051          & 0.0361          & 0.9996          & 0.9791          &  \\ \hline
hardlim                             & 0.0694          & 0.1134          & 0.0908          & 0.1517          & 0.8540          & 0.7585          &  \\ \hline
logsig                              & 0.0042          & 0.0350          & 0.0066          & 0.0552          & 0.9993          & 0.9711          &  \\ \hline
poslin                              & 0.0351          & 0.0567          & 0.0496          & 0.1039          & 0.9590          & 0.8495          &  \\ \hline
purelin                             & 0.0566          & 0.0584          & 0.0760          & 0.1004          & 0.9133          & 0.7796          &  \\ \hline
radbas                              & 0.1884          & 0.1894          & 0.2480          & 0.2522          & 0.5844          & 0.4557          &  \\ \hline
satlin                              & 0.0251          & 0.0371          & 0.0367          & 0.0547          & 0.9780          & 0.9616          &  \\ \hline
tribas                              & 0.0254          & 0.0601          & 0.0334          & 0.1089          & 0.9822          & 0.8048          &  \\ \hline
\end{tabular}
\label{tab4}%
\end{table}

\begin{figure}
	\centering
	\includegraphics[scale=0.7]{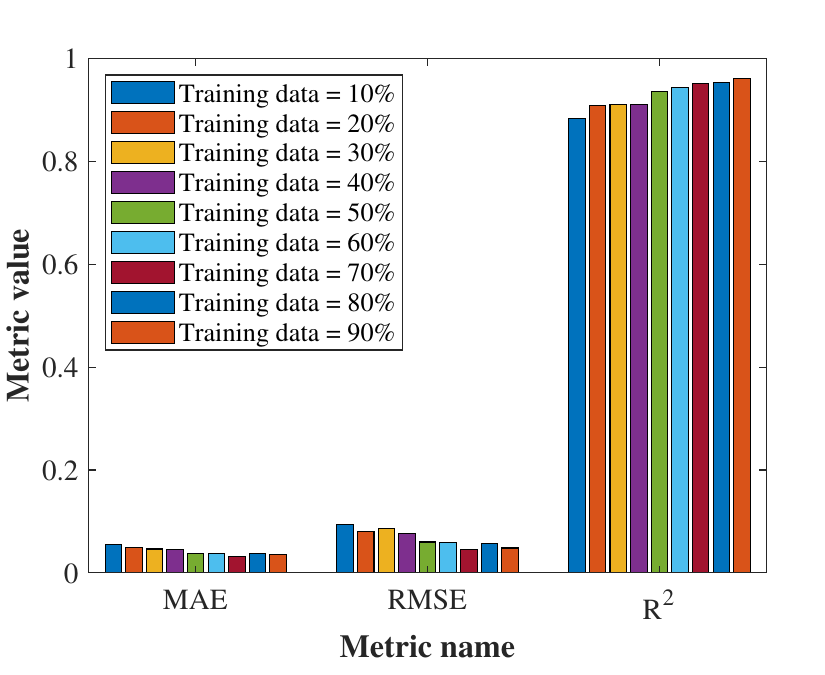}
	\caption{Comparison of the performance of the proposed BRANN model trained with different data sizes for predicting tool wear based on the NASA Ames milling data set.}
	\label{Fig.4}
\end{figure}

Subsequently, the influence of utilizing varying amounts of training data on the performance of the BRANN is investigated. Nine different ratios of training data, as illustrated in Fig.\ref{Fig.4} are considered. It can be seen that as the ratio of training data increases, the performance of the BRANN improves. Furthermore, it is observed that the BRANN trained with 70\% of the entire data set consistently outperforms other investigated options across all evaluation metrics. These results indicate that utilizing 70\% of the complete data set for training purposes is sufficient to generate a well-trained BRANN model that accurately predicts tool wear using the NASA Ames milling data set.

In addition, the performance of the proposed BRANN model is examined using different training algorithms. Table \ref{tab5} presents a comparison of the model's performance using eleven distinct training algorithms, including Bayesian regularization (trainbr) \cite{foresee1997gauss}, Gradient descent with momentum (traingdm) \cite{rehman2011effect}, Gradient descent with adaptive learning rate (traingda) \cite{xie2019local}, Gradient descent with momentum and adaptive learning rate (traingdx) \cite{yu2002backpropagation}, Levenberg-Marquardt backpropagation (trainml) \cite{sapna2012backpropagation}, Resilient backpropagation (trainrp) \cite{riedmiller1993direct}, Conjugate gradient backpropagation with Fletcher-Reeves updates (traincgf) \cite{scales1985introduction}, Conjugate gradient backpropagation with Powell-Beale restarts (traincgb) \cite{powell1977restart}, Scaled conjugate gradient backpropagation (trainscg) \cite{moller1993scaled}, Conjugate gradient backpropagation with Polak-Ribiére updates (traincgp) \cite{scales1985introduction}, and BFGS quasi-Newton backpropagation (trainbfg) \cite{cobb1982practical}. It can be seen from the table that the trainbr algorithm achieves the smallest MAE and RMSE, while obtaining the highest $R^2$ scores on both the training and test sets. These findings indicate that the trainbr algorithm surpasses all other compared algorithms in effectively training the neural network for accurate prediction of the tool wear.

\begin{table}[]
 \centering
	\caption{Comparison of the performance of the proposed BRANN model trained by different algorithms for predicting tool wear using the NASA Ames milling data set.}
\begin{tabular}{llllllll}
\hline
\multirow{2}{*}{Training algorithm}                                          & \multicolumn{2}{c}{MAE}                                 & \multicolumn{2}{c}{RMSE}                                & \multicolumn{2}{c}{$R^2$}                                   &  \\ \cline{2-8} 
                                                                             & \multicolumn{1}{c}{Training} & \multicolumn{1}{c}{Test} & \multicolumn{1}{c}{Training} & \multicolumn{1}{c}{Test} & \multicolumn{1}{c}{Training} & \multicolumn{1}{c}{Test} &  \\ \hline
\textbf{trainbr}                                 & \textbf{0.0040}              & \textbf{0.0268}          & \textbf{0.0060}              & \textbf{0.0365}          & \textbf{0.9994}              & \textbf{0.9842}          &  \\ \hline
traingdm                                 & 0.3010                       & 0.2907                   & 0.4132                       & 0.3735                   & 0.0518                       & 0.0316                   &  \\ \hline
traingda                   & 0.0468                       & 0.0855                   & 0.0578                       & 0.1188                   & 0.9451                       & 0.7815                   &  \\ \hline
traingdx       & 0.0494                       & 0.1732                   & 0.0630                       & 0.5637                   & 0.9379                       & 0.0294                   &  \\ \hline
trainml                             & 0.0319                       & 0.0653                   & 0.0394                       & 0.0742                   & 0.9762                       & 0.9669                   &  \\ \hline
trainrp                                         & 0.0417                       & 0.0547                   & 0.0555                       & 0.0671                   & 0.9531                       & 0.9225                   &  \\ \hline
traincgf & 0.0731                       & 0.0814                   & 0.0974                       & 0.1093                   & 0.8693                       & 0.7049                   &  \\ \hline
traincgb   & 0.0483                       & 0.0969                   & 0.0611                       & 0.1417                   & 0.9329                       & 0.7162                   &  \\ \hline
trainscg                       & 0.1004                       & 0.0682                   & 0.1400                       & 0.0859                   & 0.7104                       & 0.8845                   &  \\ \hline
traincgp   & 0.0214                       & 0.0565                   & 0.0276                       & 0.0701                   & 0.9879                       & 0.9203                   &  \\ \hline
trainbfg                              & 0.0294                       & 0.0894                   & 0.0373                       & 0.1357                   & 0.9769                       & 0.7641                   &  \\\hline

\end{tabular}
\label{tab5}%
\end{table}

\begin{figure*}[!htb]			
	\centering		
	\begin{subfigure}{0.45\textwidth}
		\centering
		\includegraphics[scale = 0.6]{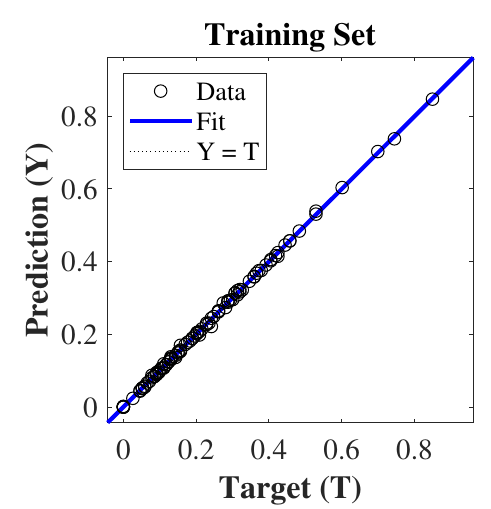}
		\caption{Regression results on the training set.}
		\label{Fig.10a}	
	\end{subfigure}
	~
	\begin{subfigure}{0.45\textwidth}
		\centering
		\includegraphics[scale = 0.6]{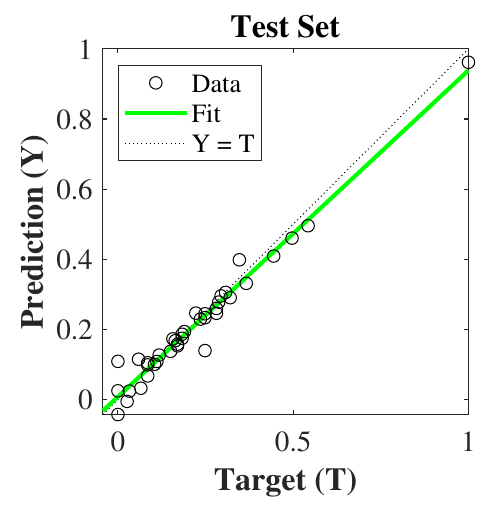}
		\caption{Regression results on the test set.}
		\label{Fig.10b}	
	\end{subfigure}%
	\caption{Regression results of the proposed BRANN model in predicting tool wear using the NASA Ames milling data set.}		
	\label{Fig.7}		
\end{figure*}

\begin{figure*}[!htb]			
	\centering		
	\includegraphics[scale = 0.7]{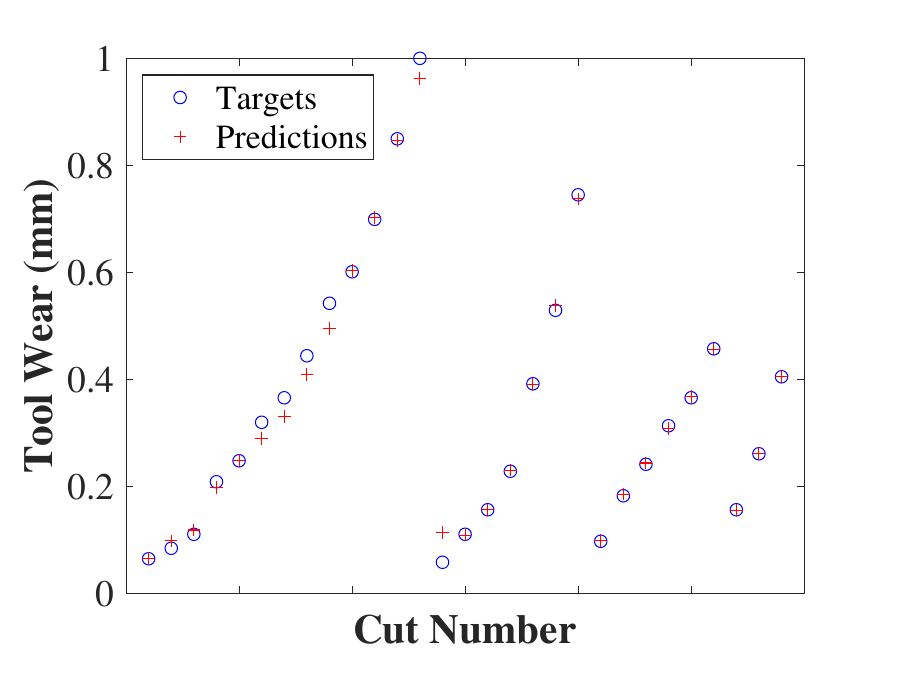}
	\caption{Tool wear prediction obtained by the proposed BRANN model using the NASA Ames milling data set.}
	\label{FigToolWear}		
\end{figure*}

Next, to validate the accuracy and reliability of the proposed BRANN, a comparative study is conducted by comparing its results with five other existing models: LR, SVR, MLP, CNN, and LSTM. The performance of these models in predicting tool wear using the NASA Ames milling data set is presented in Table \ref{tab6}. The findings demonstrate that the proposed BRANN outperforms the other five models, yielding the lowest values for MAE and RMSE. This indicates the robustness and applicability of the proposed BRANN model in accurately predicting tool wear. In addition, the regression results of the proposed BRANN model on the training and test sets are presented in Fig. \ref{Fig.7}. The depicted graphs clearly indicate that the predicted results align closely with the target values in both the training and test sets. Moreover, the tool wear prediction for different cutting conditions obtained from the proposed BRANN model are illustrated in Fig. \ref{FigToolWear}. It is seen that the proposed model can predict tool wear trends and provide accurate predictions with minimal error. These results further emphasize the strong performance and practicality of the proposed BRANN model in accurately predicting tool wear.

\begin{table}[htbp]
    \centering
	\caption{Comparison of the performance of different models in predicting the tool wear using the NASA Ames milling data set.}
	\begin{tabular}{lcc}
		\hline
		Models & MAE   & RMSE   \\
		\hline
		LR  \cite{cai2020hybrid}  & 0.1879 & 0.2146  \\
		SVR \cite{cai2020hybrid}  & 0.17  & 0.1945  \\
		MLP \cite{cai2020hybrid}  & 0.183 & 0.2168  \\
		CNN \cite{cai2020hybrid}  & 0.1835 & 0.2143  \\
		LSTM \cite{cai2020hybrid} & 0.0322 & 0.0456  \\
		\textbf{BRANN (This study)} & \textbf{0.0251} & \textbf{0.0352}  \\
		\hline
	\end{tabular}%
	\label{tab6}%
\end{table}%

\begin{figure*}[!htb]			
	\centering		
	\includegraphics[scale = 0.7]{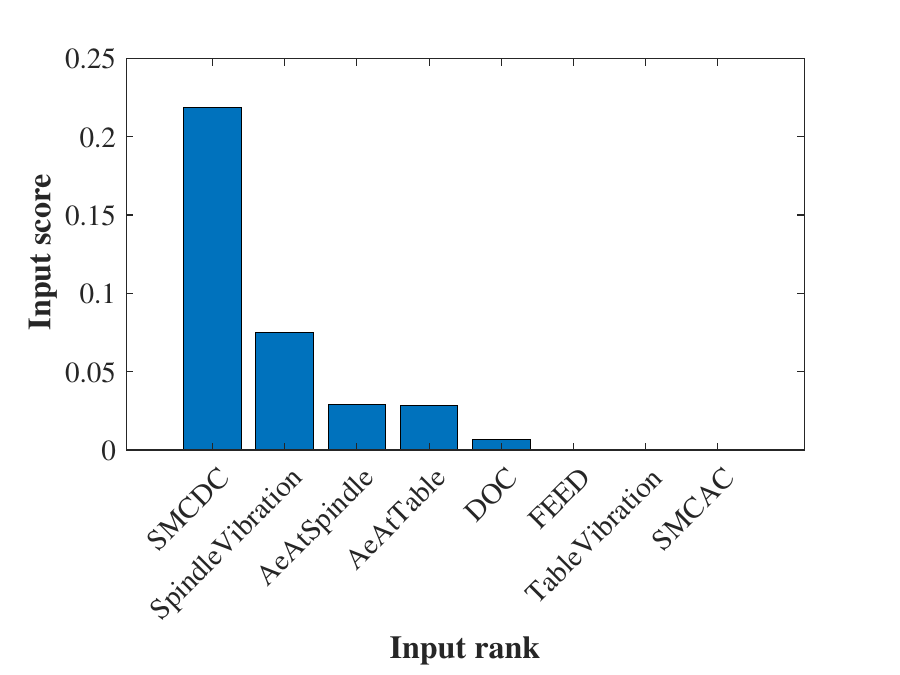}
	\caption{Important inputs in the NASA Ames milling data set for predicting the tool wear.}
	\label{Fig.5}		
\end{figure*}

Furthermore, the rank of features used as inputs of the proposed BRANN model to predict the tool wear is analyzed by a minimum redundancy maximum relevance (MRMR) algorithm \cite{ding2005minimum}. This algorithm assesses relevance and redundancy using pairwise mutual information among variables and mutual information between features and the target variable. The rank score of eight inputs, which are considered as inputs for the BRANN, is depicted in Fig. \ref{Fig.5}. Notably, four inputs, namely DOC, FEED, TableVibration and SMCAC, are identified as the least significant inputs while the SMCDC, SpindleVibration, AeAtSpindle, and AeAtTable are indentified as the most significant features. Herein, the process information (i.e., DOC and FEED) is recognized as fewer impact inputs for predicting the tool wear. This is because they do not vary for each experimental condition.

\subsection{Prediction of tool flank wear using 2010 PHM Data Challenge data set}
\label{section 4.2}
To conduct a more in-depth investigation of the performance of the proposed BRANN in predicting tool wear, the 2010 PHM Data Challenge data set is considered in this part. The case study involves force, vibration (in $x$, $y$, and $z$ directions), and AE-RMS as input parameters for the model. Meanwhile, the flank wear observed at three different flutes, namely flute 1, flute 2, and flute 3 is taken as the output of the model. Table \ref{tab7} provides a comprehensive overview of the input and output details to train the BRANN model.

\begin{table}[]
\centering
\caption{The information of inputs and outputs for training the BRANN model for predicting tool wear using the 2010 PHM Data Challenge data set.}

\begin{tabular}{llclc}
\hline
\textbf{Inputs}             & \textbf{}                                           & \textbf{} & \textbf{Ouputs}              & \multicolumn{1}{l}{\textbf{}}   \\ \hline
Monitoring   signals        & Description                                         &           & Monitoring signals           & \multicolumn{1}{l}{Description} \\ \hline
                            & Force (N) in X dimension                            &           &                              &                                 \\ \cline{2-3}
                            & Force (N) in Y   dimension                          &           &                              & \multirow{-2}{*}{Flute-1}       \\ \cline{2-3} \cline{5-5} 
\multirow{-3}{*}{Force}     & Force (N) in Z   dimension                          &           &                              &                                 \\ \cline{1-3}
                            & Vibration (g) in X dimension                        &           &                              &                                 \\ \cline{2-3}
                            & Vibration (g) in Y dimension                        &           &                              & \multirow{-3}{*}{Plute-2}       \\ \cline{2-3} \cline{5-5} 
\multirow{-3}{*}{Vibration} & {\color[HTML]{3C4043} Vibration (g) in Z dimension} &           &                              &                                 \\ \cline{1-3}
                            & AE-RMS (V)                                          &           & \multirow{-7}{*}{Flank wear} & \multirow{-2}{*}{Plute-3}       \\ \hline
\end{tabular}
\label{tab7}
\end{table}

Similarly, the impact of various factors such as input features, number of hidden units, training data size, transfer functions, and training algorithms on the predictive capability of the BRANN in relation to tool wear prediction using the 2010 PHM Data Challenge dataset is examined. Specifically, the performance of the BRANN with different numbers of hidden units is compared and presented in Table \ref{tab8}. The results demonstrate that the BRANN model with 32 hidden units consistently yields superior outcomes. This again shows that the BRANN model with 32 hidden units is a suitable choice for predicting tool wear. In addition, it is also seen from Table \ref{tab9} that the usage of the tansig as a transfer function in the BRANN gives the best performance in comparison with other investigated functions. This again indicates that the tansig function is a suitable selection as the transfer function in the proposed BRANN model for predicting the tool wear.

\begin{table}[]
\centering
\caption{Comparison of the performance of the BRANN with different numbers of hidden units in predicting tool wear using the 2010 PHM Data Challenge data set.}
\begin{tabular}{cccccccc}
\hline
\multirow{2}{*}{Number of hidden units} & \multicolumn{3}{c}{MAE}                             &           & \multicolumn{3}{c}{RMSE}                            \\ \cline{2-8} 
                                        & C1              & C4              & C6              &           & C1              & C4              & C6              \\ \hline
8                                       & 0.0115          & 0.0119          & 0.0142          &           & 0.0151          & 0.0155          & 0.0199          \\ \hline
16                                      & 0.0075          & 0.0080          & 0.0087          &           & 0.0100          & 0.0111          & 0.0121          \\ \hline
\textbf{32 }                                    & \textbf{0.0048} & \textbf{0.0052} & \textbf{0.0053} & \textbf{} & \textbf{0.0079} & \textbf{0.0073} & \textbf{0.0095} \\ \hline
64                                      & 0.0046          & 0.0048          & 0.0056          &           & 0.0163          & 0.0175          & 0.0254          \\ \hline
\end{tabular}
\label{tab8}
\end{table}

\begin{table}[]
\centering
\caption{Comparison of the performance of the BRANN with different transfer functions in predicting the tool wear using the 2010 PHM Data Challenge data set}
\begin{tabular}{cccccccc}
\hline
\multirow{2}{*}{Transfer functions} & \multicolumn{3}{c}{MAE}                             &           & \multicolumn{3}{c}{RMSE}                            \\ \cline{2-8} 
                                    & C1              & C4              & C6              &           & C1              & C4              & C6              \\ \hline
\textbf{tansig}                     & \textbf{0.0048} & \textbf{0.0052} & \textbf{0.0053} & \textbf{} & \textbf{0.0079} & \textbf{0.0073} & \textbf{0.0095} \\ \hline
compet                              & 0.0839          & 0.0905          & 0.0977          &           & 0.1100          & 0.1207          & 0.1181          \\ \hline
elliotsig                           & 0.0047          & 0.0047          & 0.0059          &           & 0.0081          & 0.0087          & 0.0145          \\ \hline
hardlim                             & 0.0453          & 0.0593          & 0.0595          &           & 0.0583          & 0.0784          & 0.0748          \\ \hline
logsig                              & 0.0051          & 0.0056          & 0.0049          &           & 0.0090          & 0.0128          & 0.0081          \\ \hline
poslin                              & 0.0133          & 0.0124          & 0.0119          &           & 0.0199          & 0.0169          & 0.0186          \\ \hline
purelin                             & 0.0289          & 0.0450          & 0.0341          &           & 0.0353          & 0.0548          & 0.0454          \\ \hline
radbas                              & 0.0058          & 0.0051          & 0.0059          &           & 0.0092          & 0.0079          & 0.0113          \\ \hline
satlin                              & 0.0101          & 0.0105          & 0.0105          &           & 0.0138          & 0.0148          & 0.0152          \\ \hline
tribas                              & 0.0128          & 0.0130          & 0.0127          &           & 0.0176          & 0.0180          & 0.0184          \\ \hline
\end{tabular}
\label{tab9}
\end{table}

Subsequently, Table \ref{tab10} presents a comparison of the performance of the BRANN trained with various data size. It is evident that as the amount of training data increases, the accuracy of the BRANN model improves. Notably, when the BRANN model is trained with 80\% and 90\% of the entire dataset, it achieves highly accurate tool wear prediction results. Additionally, it is observed that the model trained with 90\% of the entire dataset exhibits slightly higher accuracy than the one trained with 80\% of the dataset. Nevertheless, using 80\% of the data for training is sufficient to obtain a well-trained model capable of accurately predicting tool wear. Therefore, for this particular case study, 80\% of the complete dataset is utilized for training the BRANN model. 

\begin{table}[]
\centering
\caption{The comparison of the performance of the BRANN trained with different training data ratios for predicting the tool wear using the the 2010 PHM Data Challenge data set.}
\begin{tabular}{cccccccc}
\hline
\multirow{2}{*}{Training data (\%)} & \multicolumn{3}{c}{MAE}                             &           & \multicolumn{3}{c}{RMSE}                            \\ \cline{2-8} 
                                    & C1              & C4              & C6              &           & C1              & C4              & C6              \\ \hline
10                                  & 0.0229          & 0.0162          & 0.0301          &           & 0.0462          & 0.0162          & 0.0592          \\ \hline
20                                  & 0.0207          & 0.0137          & 0.0260          &           & 0.0694          & 0.0258          & 0.0515          \\ \hline
30                                  & 0.0135          & 0.0109          & 0.0288          &           & 0.0281          & 0.0198          & 0.0745          \\ \hline
40                                  & 0.0084          & 0.0085          & 0.0116          &           & 0.0169          & 0.0173          & 0.0288          \\ \hline
50                                  & 0.0057          & 0.0071          & 0.0089          &           & 0.0093          & 0.0151          & 0.0218          \\ \hline
60                                  & 0.0057          & 0.0071          & 0.0089          &           & 0.0093          & 0.0151          & 0.0218          \\ \hline
70                                  & 0.0064          & 0.0051          & 0.0070          &           & 0.0148          & 0.0072          & 0.0151          \\ \hline
\textbf{80}                         & \textbf{0.0048} & \textbf{0.0052} & \textbf{0.0053} & \textbf{} & \textbf{0.0079} & \textbf{0.0073} & \textbf{0.0095} \\ \hline
90                                  & 0.0049          & 0.0047          & 0.0045          &           & 0.0076          & 0.0061          & 0.0066          \\ \hline
\end{tabular}
\label{tab10}
\end{table}

\begin{figure}
	\centering
	\includegraphics[scale=0.7]{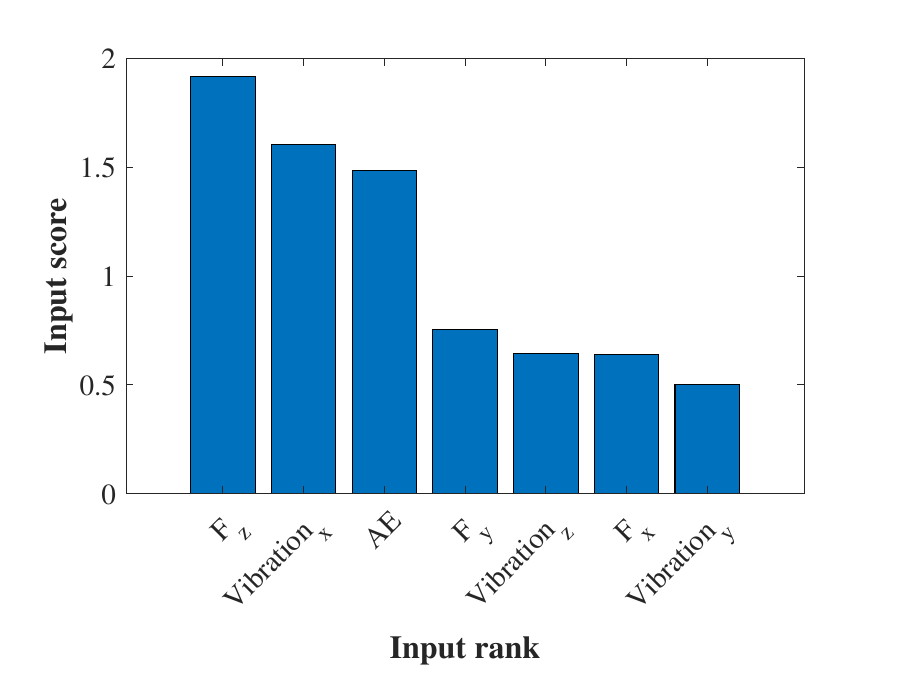}
	\caption{Important inputs in the 2010 PHM Data Challenge data set for predicting the tool wear.}
\label{Fig.9}
\end{figure}

Additionally, Table \ref{tab11} presents a comparison of the BRANN's performance when trained using various training algorithms. Once again, it is evident that the trainbr algorithm outperforms all the other algorithms in effectively training the neural network to achieve precise tool wear prediction. Furthermore, Fig. \ref{Fig.9} presents an analysis of the significance of each individual inputs in the 2010 PHM Data Challenge dataset for predicting the tool wear. It can be seen that all the features have a substantial influence on tool wear and are crucial inputs for training the BRANN model. In addition, the figure also indicates that the force signal in the $z$ direction is the most significant factor for predicting the tool wear, whereas the vibration signal in the $y$ direction has the least significant impact among seven considered inputs.

\begin{table}[]
\centering
\small

\caption{Comparison of the performance of the BRANN trained with different algorithms for predicting tool wear using the 2010 PHM Data Challenge data set.}
\begin{tabular}{cccccccc}
\hline
\multirow{2}{*}{Training algorithm} & \multicolumn{3}{c}{MAE}                             &           & \multicolumn{3}{c}{RMSE}                            \\ \cline{2-8} 
                                    & C1              & C4              & C6              &           & C1              & C4              & C6              \\ \hline
\textbf{trainbr}                    & \textbf{0.0048} & \textbf{0.0052} & \textbf{0.0053} & \textbf{} & \textbf{0.0079} & \textbf{0.0073} & \textbf{0.0095} \\ \hline
traingdm                            & 0.0954          & 0.0966          & 0.1220          &           & 0.1427          & 0.1354          & 0.1630          \\ \hline
traingda                            & 0.0684          & 0.0781          & 0.0840          &           & 0.0989          & 0.1105          & 0.1087          \\ \hline
traingdx                            & 0.0485          & 0.0516          & 0.0540          &           & 0.0640          & 0.0667          & 0.0715          \\ \hline
trainml                             & 0.0125          & 0.0114          & 0.0133          &           & 0.0165          & 0.0149          & 0.0200          \\ \hline
trainrp                             & 0.0331          & 0.0357          & 0.0345          &           & 0.0420          & 0.0441          & 0.0451          \\ \hline
traincgf                            & 0.0227          & 0.0252          & 0.0209          &           & 0.0289          & 0.0315          & 0.0279          \\ \hline
traincgb                            & 0.0319          & 0.0410          & 0.0309          &           & 0.0396          & 0.0498          & 0.0404          \\ \hline
trainscg                            & 0.0286          & 0.0280          & 0.0303          &           & 0.0360          & 0.0355          & 0.0403          \\ \hline
traincgp                            & 0.0462          & 0.0463          & 0.0485          &           & 0.0604          & 0.0575          & 0.0675          \\ \hline
trainbfg                            & 0.0299          & 0.0351          & 0.0285          &           & 0.0378          & 0.0434          & 0.0376          \\ \hline
\end{tabular}
\label{tab11}
\end{table}

\begin{figure*}[!htb]			
	\centering		
	\begin{subfigure}{0.45\textwidth}
		\centering
		\includegraphics[scale = 0.7]{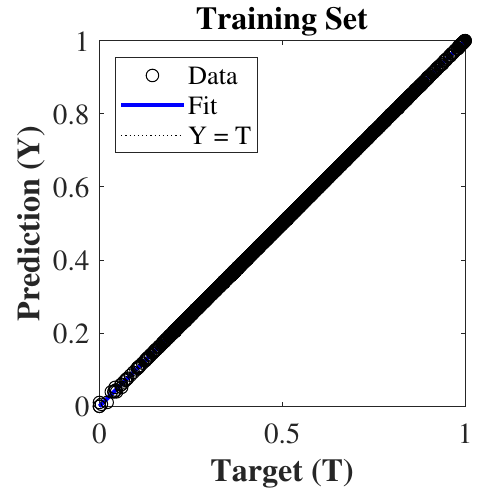}
		\caption{Regression results on the training set.}
		\label{Fig.10a}	
	\end{subfigure}
	~
	\begin{subfigure}{0.45\textwidth}
		\centering
		\includegraphics[scale = 0.7]{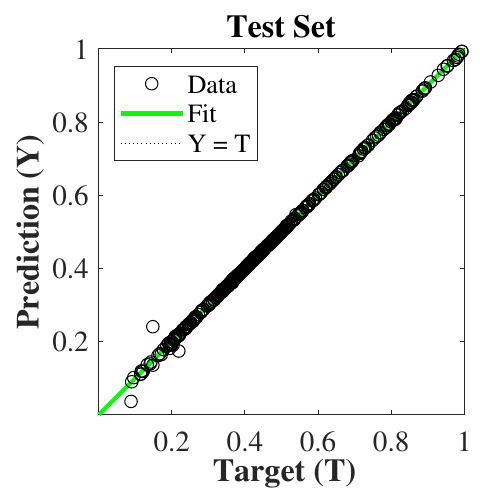}
		\caption{Regression results on the test set.}
		\label{Fig.10b}	
	\end{subfigure}%
	\caption{Regression results of the proposed BRANN model for predicting tool wear using the 2010 PHM Data Challenge data set.}		
	\label{Fig.13}		
\end{figure*}

\begin{figure*}[!htb]			
	\centering		
	\begin{subfigure}{0.3\textwidth}
		\centering
		\includegraphics[scale = 0.35]{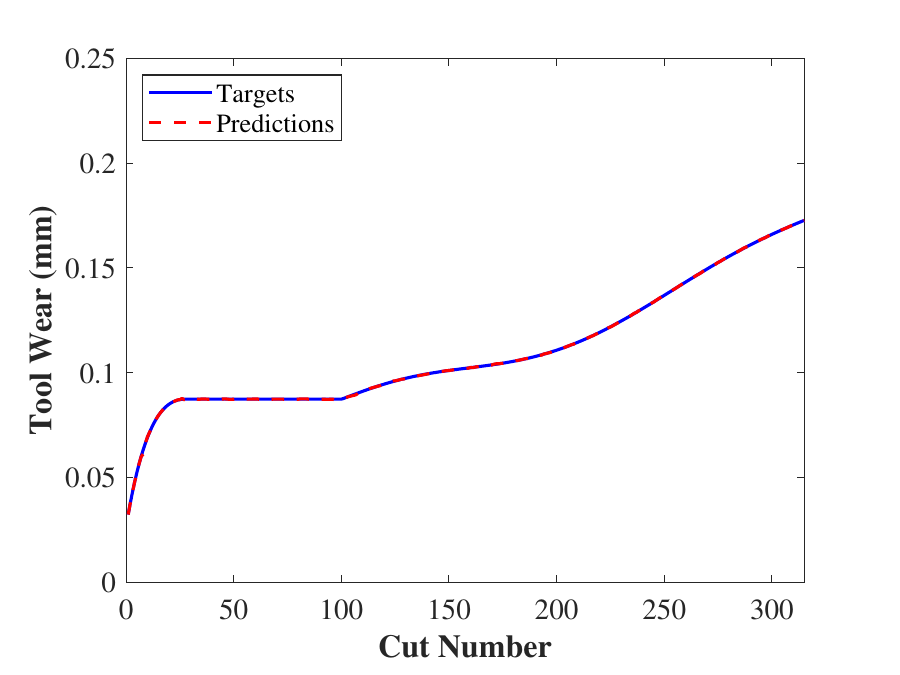}
		\caption{Tool wear prediction for C1.}
	\end{subfigure}
	~
	\begin{subfigure}{0.3\textwidth}
		\centering
		\includegraphics[scale = 0.35]{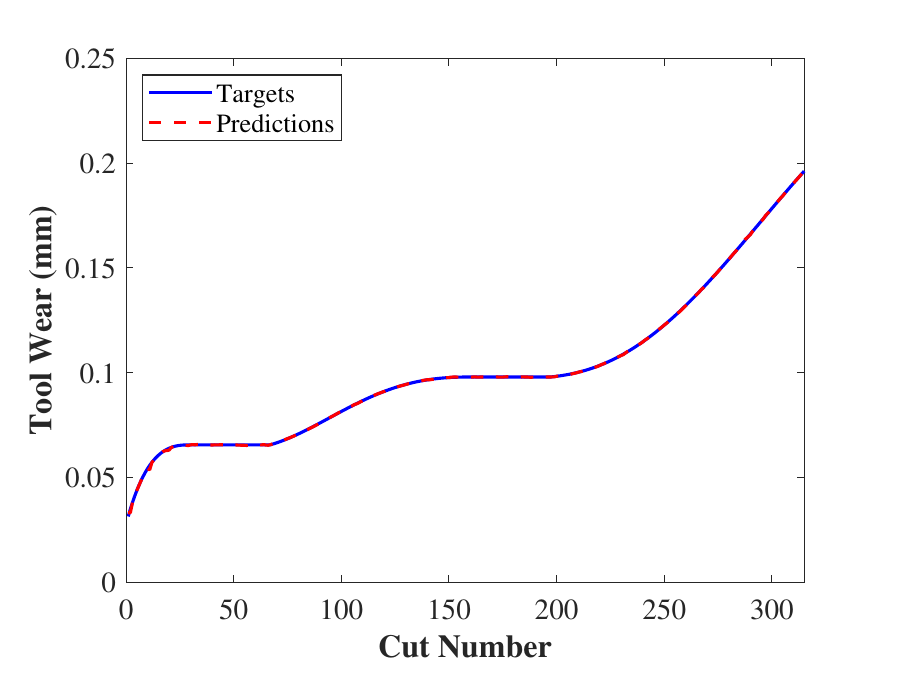}
		\caption{Tool wear prediction for C4.}
	\end{subfigure}%
       ~
	\begin{subfigure}{0.3\textwidth}
		\centering
		\includegraphics[scale = 0.35]{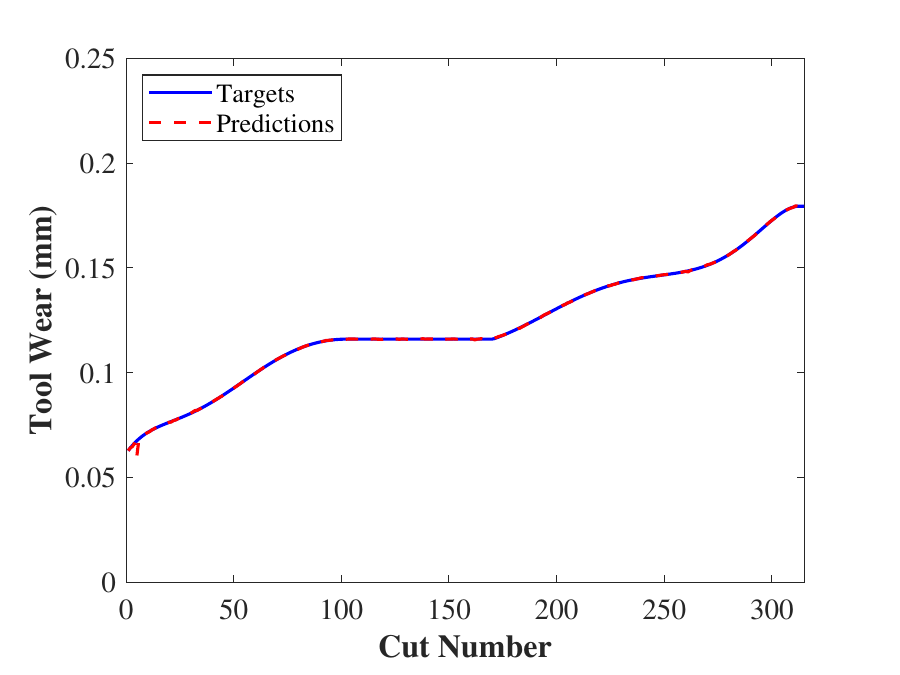}
		\caption{Tool wear prediction for C6.}
	\end{subfigure}%
	\caption{Tool wear predicting results obtained by the proposed BRANN model using the 2010 PHM Data Challenge data set.}	
	\label{FigToolWearPredictionPHM}		
\end{figure*}

Moreover, in order to further demonstrate the precision and robustness of the proposed BRANN model in forecasting the tool wear, its effectiveness is evaluated by comparing it with five other existing models: LR, SVR, MLP, CNN, and LSTM. The comparative outcomes are presented in Table \ref{tab12}. Evidently, the table reveals that the proposed BRANN consistently achieved the lowest MAE and RMSE values across all three scenarios (C1, C4, and C6) when compared to the other existing machine learning and deep learning models. This clearly demonstrates that the proposed BRANN model surpasses LR, SVR, MLP, CNN, and LSTM models in accurately predicting tool wear. Furthermore, Fig.\ref{Fig.13} shows the regression results of the proposed BRANN model on the training and test sets in predicting tool wear using the 2010 PHM Data Challenge data set. It is seen that the predicted results closely align with the experimental results (i.e., targets) in both training set and test set. In addition, it can be seen from Fig. \ref{FigToolWearPredictionPHM} that the proposed BRANN model can accurately predict the tool wear evolution and estimate the tool wear values with a small error in comparison with the target values obtained from the experiment. These results once again highlight the precision and applicability of the proposed BRANN model in accurately predicting tool wear.

\begin{table}[]
\centering
\small
\caption{Comparison of the accuracy of different models in predicting the tool wear using the 2010 PHM Data Challenge data set.}
\begin{tabular}{lccccccc}
\hline
\multirow{2}{*}{Models}      & \multicolumn{3}{c}{MAE}                                            &                      & \multicolumn{3}{c}{RMSE}                                           \\ \cline{2-8} 
                             & C1                   & C4                   & C6                   &                      & C1                   & C4                   & C6                   \\ \hline
LR \cite{cai2020hybrid}                          & 0.0337               & 0.0164               & 0.0668               &                      & 0.0501               & 0.0189               & 0.0904               \\ \hline
SVR  \cite{cai2020hybrid}                        & 0.0098               & 0.0262               & 0.0273               &                      & 0.0144               & 0.0293               & 0.0389               \\ \hline
MLP  \cite{cai2020hybrid}                        & 0.0251               & 0.0336               & 0.0268               &                      & 0.0288               & 0.0398               & 0.0336               \\ \hline
CNN  \cite{cai2020hybrid}                        & 0.0254               & 0.0391               & 0.0383               &                      & 0.0293               & 0.0436               & 0.0553               \\ \hline
LSTM \cite{cai2020hybrid}                        & 0.0085               & 0.0085               & 0.0146               &                      & 0.0114               & 0.0117               & 0.0212               \\ \hline
\textbf{BRANN (This   study)} & \textbf{0.0004}      & \textbf{0.0006}      & \textbf{0.0007}      & \textbf{}            & \textbf{0.0008}      & \textbf{0.0038}      & \textbf{0.0011}      \\ \hline
                             & \multicolumn{1}{l}{} & \multicolumn{1}{l}{} & \multicolumn{1}{l}{} & \multicolumn{1}{l}{} & \multicolumn{1}{l}{} & \multicolumn{1}{l}{} & \multicolumn{1}{l}{}
\end{tabular}
\label{tab12}
\end{table}

\subsection{Prediction of the tool wear using the NUAA Ideahouse tool wear data set}
For further investigation the performance of the BRANN in predicting the tool wear, a dataset of NUAA Ideahouse including thirteen cutting conditions is investigated in this part. Herein, the monitoring signals for force, vibration and opc are considered as inputs of the model while the flank wear at four different edges of the tool are taken into account as the outputs of the model. The detailed description of the inputs and outputs of the model is presented in Table \ref{tab13}. Similar to subsection \ref{section 4.2}, the BRANN is trained by the trainbr algorithm using 80\% of the entire data. The tansig is also chosen as a transfer function of the BRANN model. The hidden units of the BRANN model is set to 16 in this case.

\begin{table}[]
\centering
\small
\caption{The information of inputs and outputs for training the BRANN model for predicting tool wear using the NUAA Ideahouse tool wear dataset.}
\begin{tabular}{ccccc}
\hline
\multicolumn{2}{c}{Inputs}                             &  & \multicolumn{2}{c}{Ouputs}                            \\ \hline
Monitoring   signals       & Description               &  & Monitoring signals          & Description             \\ \hline
\multirow{4}{*}{Force}     & Axial Force/N             &  & \multirow{8}{*}{Flank wear} & \multirow{2}{*}{Edge-1} \\ \cline{2-3}
                           & Bending Moment   of X/N.m &  &                             &                         \\ \cline{2-3} \cline{5-5} 
                           & Bending Moment   of Y/N.m &  &                             & \multirow{2}{*}{Edge-2} \\ \cline{2-3}
                           & Torsion of Z/N.m          &  &                             &                         \\ \cline{1-3} \cline{5-5} 
\multirow{2}{*}{Vibration} & Channel 1                 &  &                             & \multirow{2}{*}{Edge-3} \\ \cline{2-3}
                           & Channel 2                 &  &                             &                         \\ \cline{1-3} \cline{5-5} 
\multirow{2}{*}{Opc}       & Spindle power             &  &                             & \multirow{2}{*}{Edge-4} \\ \cline{2-3}
                           & Spindle current           &  &                             &                         \\ \hline
\end{tabular}
\label{tab13}
\end{table}

\begin{figure*}[!htb]			
	\centering		
	\begin{subfigure}{0.45\textwidth}
		\centering
		\includegraphics[scale = 0.7]{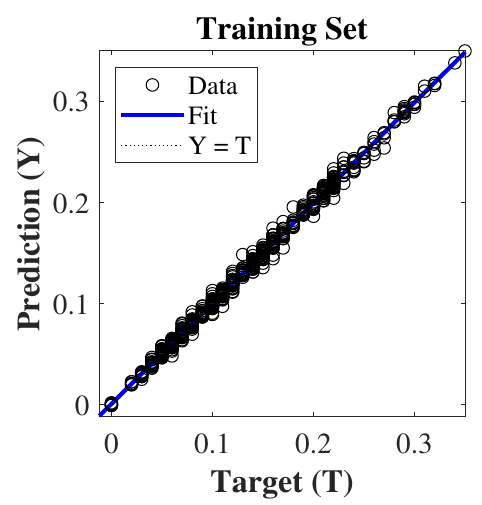}
		\caption{Regression results on the training set.}
		\label{Fig.10a}	
	\end{subfigure}
	~
	\begin{subfigure}{0.45\textwidth}
		\centering
		\includegraphics[scale = 0.7]{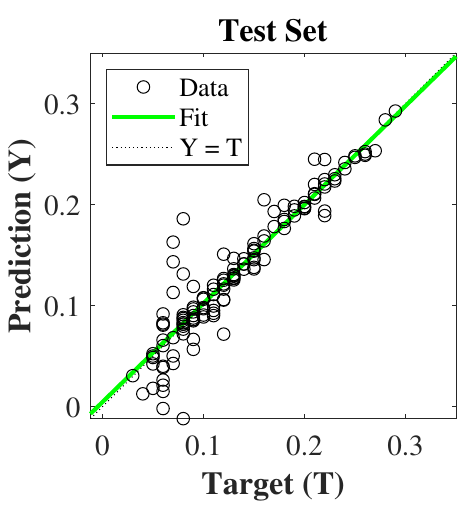}
		\caption{Regression results on the test set.}
		\label{Fig.10b}	
	\end{subfigure}%
	\caption{Regression results of the proposed BRANN model for predicting the tool wear using the NUAA Ideahouse dataset.}		
	\label{Fig.19}		
\end{figure*}

Table \ref{tab14} illustrates a comparison between the prediction accuracy of the proposed BRANN model and the existing MIFS model \cite{liu2021meta} across various cutting conditions. The results clearly indicate that the BRANN model outperforms the MIFS model, exhibiting smaller MAE and RMSE in all cases. This signifies the superior accuracy of the BRANN model in predicting tool wear when compared to the MIFS model. Consequently, these findings further validate the efficacy and superiority of the proposed BRANN model over the existing approach in tool wear prediction. In addition, Fig.\ref{Fig.19} shows the regression results of the proposed BRANN model on training and testing sets. The result again shows the good performance of the proposed BRANN in both training and test sets for predicting the tool wear. 

\begin{table}[]
\centering
\caption{Comparison of the accuracy of different models in predicting tool wear using the NUAA Ideahouse tool wear dataset.}
\begin{tabular}{cccc}
\hline
Cutting   Condition & Model                    & MAE             & RMSE            \\ \hline
\multirow{2}{*}{10} & \multicolumn{1}{l}{MIFS \cite{liu2021meta}} & 0.027           & 0.031           \\ \cline{2-4} 
                    & \textbf{BRANN}           & \textbf{0.0061} & \textbf{0.016} \\ \hline
\multirow{2}{*}{11} & \multicolumn{1}{l}{MIFS \cite{liu2021meta}} & 0.032           & 0.037           \\ \cline{2-4} 
                    & \textbf{BRANN}           & \textbf{0.00066} & \textbf{0.00084} \\ \hline
\multirow{2}{*}{12} & \multicolumn{1}{l}{MIFS \cite{liu2021meta}} & 0.063           & 0.027           \\ \cline{2-4} 
                    & \textbf{BRANN}           & \textbf{0.00080} & \textbf{0.0012} \\ \hline
\multirow{2}{*}{13} & \multicolumn{1}{l}{MIFS \cite{liu2021meta}} & 0.050           & 0.032           \\ \cline{2-4} 
                    & \textbf{BRANN}           & \textbf{0.0012} & \textbf{0.0020} \\ \hline
\end{tabular}
\label{tab14}
\end{table}

\begin{figure*}[!htb]			
	\centering		
	\includegraphics[scale = 0.6]{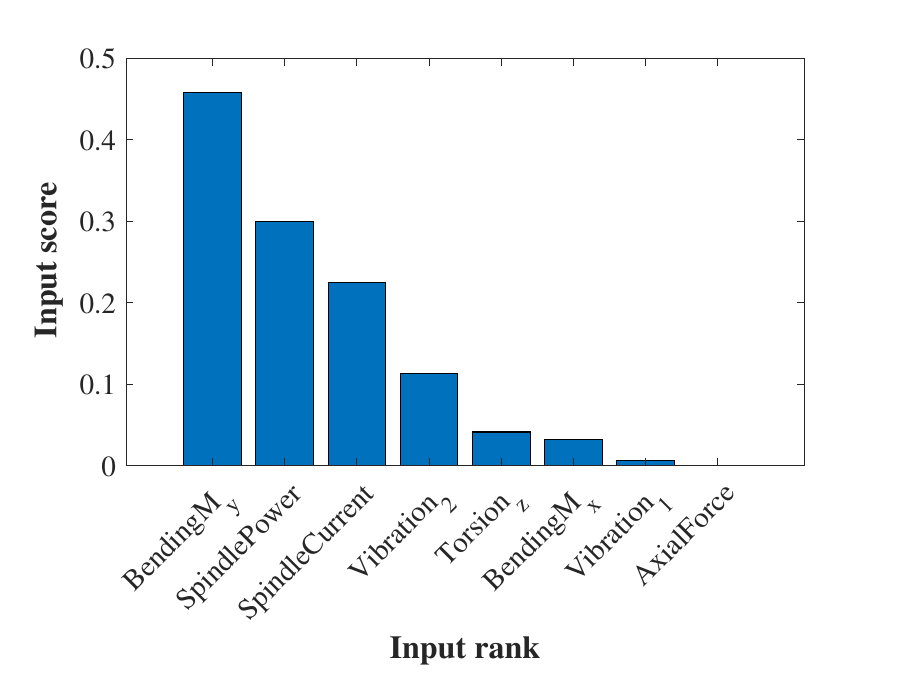}
	\caption{Important inputs in the NUAA Ideahouse dataset for predicting the tool wear.}
	\label{FigInputRankNUAA}		
\end{figure*}

Furthermore, the important inputs in the NUAA Ideahouse dataset for predicting the tool wear are analyzed and presented in Fig. \ref{FigInputRankNUAA}. It can be seen from the figure that almost inputs have certain contributions and have significant impacts on the performance of the BRANN model for predicting the tool wear. In addition, it is interesting to observe that the bending moment in $y$ direction has the highest weight (weight = 0.46) among the eight inputs, followed by  spindle power (weight = 0.29), spindle current (weight = 0.23), vibration signal at channel 2 (weight = 0.11), torsion signal in $z$ direction (weight = 0.04), bending moment in $x$ direction (weight = 0.03), vibration signal at channel 1 (weight =  0.0067) and axial force (weight = 0.00054). 

\subsection{Analysis and validation of tool condition}

As discussed earlier, the end milling was performed at different combinations of process parameters taking Ti6Al4V as workpiece material. The Ti6Al4V is considered difficult-to-cut material due to its chemical affinity and low thermal conductivity \cite{airao2023comparative}. When the Ti6Al4V undergone the machining, it accelerate the tool wear, producing higher cutting forces. The experiments performed were in dry conditions, which is favourable conditions to adhere the workpiece with the tool under a high pressure and temperature. It results in a formation of built-up edges. When this built-up edges escape away by the chip, it brings some tool material with it, producing a tool wear. Since, the milling is an intermittent cutting process, this adhesion and removal of the built-up edge enhances the chance of tool wear. In addition to that, the cyclic force acting on the cutting edge of the tool softens the tool, enabling the chance of plastic deformation \cite{airao2022machinability}. Here, Systematic experiments were executed at different cutting speeds and feed rates. The tool wear is not measured but once the tool was found broken than experiment was stopped at a particular combination of process parameter. Table \ref{tab15} shows the tool condition for each combination of process parameter. It is observed that for sample 1 where the cutting speed used was 50 m/min, no tool breakage was found. For the samples 2, 3 and 4, where the cutting speed was 60 m/min and feed rates were 45 and 50 micron/tooth. It is noted that the tool was broken when feed rate increased. Similarly, for samples 5-11, the experiments were conducted at cutting speed of 75 m/min and feed rates varying from 22 micron/tooth to 40 micron/tooth. The tool was worn out at the feed rate of 40 micro/tooth (sample 11). Likewise, the experiments at 80 m/min and 20.6-25 micron/tooth show the tool breakage for sample 14. In all the above cases, keeping cutting speed constant and by increasing the feed rate, the uncut chip thickness increases, enhancing pressure acting on the tool. Moreover, the tool in the intermittent cutting action has to be prone to remove more volume of material at higher feed rate, thus the thrust force acting on the tool becomes higher. It leads to deform the tool plastically by reducing its hardness. On the other hand, as the cutting speed increases, the corresponding feed rate at which the tool breakage found decrease. It is due to the fact that at a higher cutting speed, the heat generated at deformation zone raises due to severe plastic deformation. The thermal conductivity of the Ti6Al4V limits the heat conduction by the chip and hence get accumulated near the cutting edge of the tool. This heat is sufficient to lower the strength of the tool and thus enhances the tool wear even at low feed rate \cite{airao2022novel}.

\begin{figure*}[!htb]			
	\centering		
	\includegraphics[scale = 0.7]{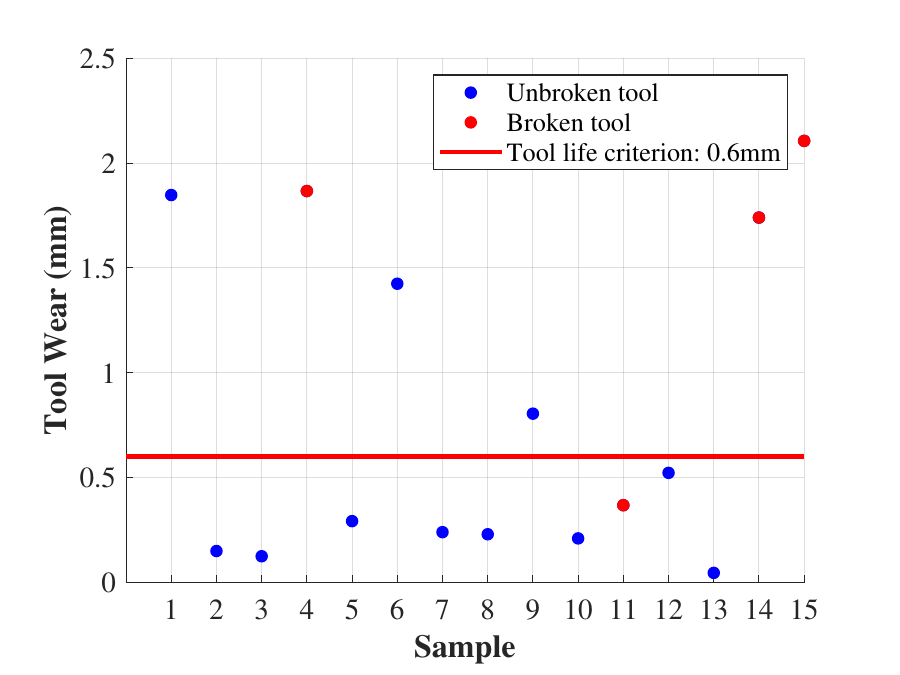}
	\caption{Tool wear prediction obtained by the proposed BRANN using the inhouse performed end milling of Ti6Al4V dataset.}
	\label{FigToolWearGerman}		
\end{figure*}

Herein, we further validate the generalization of the BRANN model by applying it to an inhouse performed end milling of Ti6Al4V dataset for classifying tool conditions. Since the tool wear values are unavailable, the tool wear classification is made by training different data sets. Specifically, the BRANN model is trained on a combined dataset comprising three datasets: the NASA Ames milling dataset, the 2010 PHM Data Challenge dataset, and the NUAA Ideahouse tool wear dataset. By combining the three datasets, the input parameters of the model include all input information of the three datasets that are given in  Table \ref{tab3}, \ref{tab7} and Table \ref{tab13}, respectively. After training, the model is then tested on an in-house dataset of the end milling of Ti6Al4V, to predict tool wear within the classification dataset. Subsequently, based on the predicted tool wear values, we classify the tool conditions into two categories: tool breakage and unbroken tool. To determine these classifications, we rely on the standard recommended value for defining a tool life criterion based on ISO 3685:1993 \cite{rizal2017cutting}. In this experimentation, the tool wear was not measured progressively, instead, the tool breakage was considered to discard the tool. Since, the tool condition classification is done based on the tool breakage, the criteria decided is the maximum value tool flank wear. i.e., 0.6 mm. The trained BRANN model predicts the tool flank wear by using four sensor input signals, including forces in the $x$, $y$, and $z$ directions, as well as the acoustic emission. A label tool condition classification data set containing 15 samples divided into two tool condition classes is presented in Table \ref{tab15}. The corresponding predicted tool flank wear results for each tool condition classification are illustrated in Fig. \ref{FigToolWearGerman}. It can be observed that the proposed model predicts tool flank wear values greater than 0.6 mm for 3 samples in the tool breakage class (except sample 11) and predicts tool flank wear values smaller than 0.6 mm for 8 out of 11 samples in the unbroken tool class. Only 3 samples (see samples 1, 6, and 9) from the unbroken tool class have predicted tool wear values greater than 0.6 mm. Thus, based on the tool life criterion, it can  classify the tool conditions based on the tool wear prediction. To be more specific, the model achieves a 75\% accuracy in classifying tool breakage conditions and a 72.73\% accuracy in classifying unbroken tool conditions. In general, the proposed model demonstrates a 73.33\% accuracy in classifying tool conditions based on tool wear predictions, even when applied to a previously unseen data. These results demonstrate the applicability and generalization of the proposed BRANN model in evaluating the tool conditions. 

Furthermore, it should be emphasized here that the BRANN model presented in this context is trained using a combination of 23 input parameters drawn from the three mentioned datasets. However, the in-house conducted end milling of Ti6Al4V dataset contains only four sensor signals (i.e., forces in the $x$, $y$, $z$ directions, and acoustic emission) that align with the trained input parameters of the combined datasets. Consequently, when predicting tool wear results in Fig. \ref{FigToolWearGerman}, the proposed BRANN model relies only on these four inputs out of the 23 available. Remarkably, despite the limited input information, the model achieves a 73.33\% accuracy in classifying tool conditions. This observation highlights that the proposed model can effectively predict tool wear with commendable accuracy even when operating with restricted input data.

\begin{table}[]
\centering
\caption{The tool condition for each combination of process parameter in the inhouse performed end milling of Ti6Al4V dataset.}
\begin{tabular}{cccccl}
\hline
Sample                                            & Cutting speed. vc {[}m/min{]}                     & Feed rate fz {[}µm/tooth{]}                       & \multicolumn{1}{l}{No of cycles}                 & Tool breakage                                      &  \\ \hline
\cellcolor[HTML]{FFFFFF}1                         & \cellcolor[HTML]{FFFFFF}50                        & \cellcolor[HTML]{FFFFFF}30                        & \cellcolor[HTML]{FFFFFF}6                        & \cellcolor[HTML]{FFFFFF}NO                         &  \\ \hline
\cellcolor[HTML]{FFFFFF}2                         & \cellcolor[HTML]{FFFFFF}60                        & \cellcolor[HTML]{FFFFFF}45                        & \cellcolor[HTML]{FFFFFF}6                        & \cellcolor[HTML]{FFFFFF}NO                         &  \\ \hline
\cellcolor[HTML]{FFFFFF}3                         & \cellcolor[HTML]{FFFFFF}60                        & \cellcolor[HTML]{FFFFFF}50                        & \cellcolor[HTML]{FFFFFF}6                        & \cellcolor[HTML]{FFFFFF}NO                         &  \\ \hline
\cellcolor[HTML]{FFFFFF}{\color[HTML]{FF0000} 4}  & \cellcolor[HTML]{FFFFFF}{\color[HTML]{FF0000} 60} & \cellcolor[HTML]{FFFFFF}{\color[HTML]{FF0000} 50} & \cellcolor[HTML]{FFFFFF}{\color[HTML]{FF0000} 4} & \cellcolor[HTML]{FFFFFF}{\color[HTML]{FF0000} YES} &  \\ \hline
\cellcolor[HTML]{FFFFFF}5                         & \cellcolor[HTML]{FFFFFF}75                        & \cellcolor[HTML]{FFFFFF}22                        & \cellcolor[HTML]{FFFFFF}6                        & \cellcolor[HTML]{FFFFFF}NO                         &  \\ \hline
\cellcolor[HTML]{FFFFFF}6                         & \cellcolor[HTML]{FFFFFF}75                        & \cellcolor[HTML]{FFFFFF}25                        & \cellcolor[HTML]{FFFFFF}6                        & \cellcolor[HTML]{FFFFFF}NO                         &  \\ \hline
\cellcolor[HTML]{FFFFFF}7                         & \cellcolor[HTML]{FFFFFF}75                        & \cellcolor[HTML]{FFFFFF}28                        & \cellcolor[HTML]{FFFFFF}6                        & \cellcolor[HTML]{FFFFFF}NO                         &  \\ \hline
\cellcolor[HTML]{FFFFFF}8                         & \cellcolor[HTML]{FFFFFF}75                        & \cellcolor[HTML]{FFFFFF}31                        & \cellcolor[HTML]{FFFFFF}6                        & \cellcolor[HTML]{FFFFFF}NO                         &  \\ \hline
\cellcolor[HTML]{FFFFFF}9                         & \cellcolor[HTML]{FFFFFF}75                        & \cellcolor[HTML]{FFFFFF}34                        & \cellcolor[HTML]{FFFFFF}6                        & \cellcolor[HTML]{FFFFFF}NO                         &  \\ \hline
\cellcolor[HTML]{FFFFFF}10                        & \cellcolor[HTML]{FFFFFF}75                        & \cellcolor[HTML]{FFFFFF}40                        & \cellcolor[HTML]{FFFFFF}6                        & \cellcolor[HTML]{FFFFFF}NO                         &  \\ \hline
\cellcolor[HTML]{FFFFFF}{\color[HTML]{FF0000} 11} & \cellcolor[HTML]{FFFFFF}{\color[HTML]{FF0000} 75} & \cellcolor[HTML]{FFFFFF}{\color[HTML]{FF0000} 40} & \cellcolor[HTML]{FFFFFF}{\color[HTML]{FF0000} 6} & \cellcolor[HTML]{FFFFFF}{\color[HTML]{FF0000} YES} &  \\ \hline
\cellcolor[HTML]{FFFFFF}12                        & \cellcolor[HTML]{FFFFFF}80                        & \cellcolor[HTML]{FFFFFF}20.6                      & \cellcolor[HTML]{FFFFFF}6                        & \cellcolor[HTML]{FFFFFF}NO                         &  \\ \hline
\cellcolor[HTML]{FFFFFF}13                        & \cellcolor[HTML]{FFFFFF}80                        & \cellcolor[HTML]{FFFFFF}25                        & \cellcolor[HTML]{FFFFFF}6                        & \cellcolor[HTML]{FFFFFF}NO                         &  \\ \hline
\cellcolor[HTML]{FFFFFF}{\color[HTML]{FF0000} 14} & \cellcolor[HTML]{FFFFFF}{\color[HTML]{FF0000} 80} & \cellcolor[HTML]{FFFFFF}{\color[HTML]{FF0000} 25} & \cellcolor[HTML]{FFFFFF}{\color[HTML]{FF0000} 6} & \cellcolor[HTML]{FFFFFF}{\color[HTML]{FF0000} YES} &  \\ \hline
\cellcolor[HTML]{FFFFFF}{\color[HTML]{FF0000} 15} & \cellcolor[HTML]{FFFFFF}{\color[HTML]{FF0000} 95} & \cellcolor[HTML]{FFFFFF}{\color[HTML]{FF0000} 17} & \cellcolor[HTML]{FFFFFF}{\color[HTML]{FF0000} 5} & \cellcolor[HTML]{FFFFFF}{\color[HTML]{FF0000} YES} &  \\ \hline
\end{tabular}
\label{tab15}
\end{table}

\clearpage
\section{Conclusions}
An effective BRANN model has been developed and successfully applied for accurately predicting tool wear. The performance and applicability of the proposed BRANN model have been demonstrated through four distinct experimental datasets including the NASA Ames milling dataset, the 2010 PHM Data Challenge dataset, the NUAA Ideahouse tool wear dataset, and the dataset of inhouse performed end milling of Ti6Al4V. The outcomes of the BRANN model are compared with those of other existing models to illustrate the accuracy and reliability of the proposed model. In addition, the impact of input features, training data size, hidden units, training algorithms, and transfer functions on the performance of the BRANN model has been examined. Based on the presented theory and experimental results, some remarkable conclusions are drawn as follows:
\begin{enumerate}
    \item The proposed BRANN model successfully avoids overfitting, leading to highly accurate tool wear predictions on both training and test sets.
    \item Under the same training dataset, the proposed BRANN outperforms existing models including LR, SVR, MLP, CNN, LSTM, and MIFS in predicting tool wear in all evaluation metrics;
    \item The input features, training data size, hidden units, training algorithms, and transfer functions have a significant impact on the performance of the BRANN model in predicting the tool wear;
    \item The proposed BRANN model has demonstrated its applicability and generalization in predicting tool wear not only under the same cutting conditions but also under different cutting conditions.
    \item By training the proposed BRANN model on a combined dataset that includes three datasets: the NASA Ames milling dataset, the 2010 PHM Data Challenge dataset, and the NUAA Ideahouse tool wear dataset, the proposed model is able to achieve a 73.33\% accuracy in classifying tool conditions when applied to the in-house performed end milling of the Ti6Al4V dataset, despite the limited input information.
\end{enumerate}

\section*{Acknowledgments}

RA and PK acknowledge funding from Villum Foundation synergy project (grant CutMore). 

\printbibliography
\end{document}

%% file: packages.tex
\usepackage[utf8]{inputenc}
\usepackage{authblk}
\usepackage{setspace}
\usepackage[margin=1.25in]{geometry}
\usepackage{graphicx}
\graphicspath{ {./figures/} }
\usepackage{subcaption}
\usepackage{amsmath}
\usepackage{lineno}
\usepackage{multirow}
\usepackage{rotating}
\usepackage{pdflscape}
\usepackage[table,xcdraw]{xcolor}
\linenumbers
\usepackage{comment}
\usepackage{xcolor} 

%% file: biblio.tex
\usepackage[style=nejm, 
citestyle=numeric-comp,
sorting=none]{biblatex}

%% file: authors.tex
\author[1]{Tam T. Truong}
\author[2]{Jay Airao}
\author[1]{Panagiotis Karras}
\author[3]{Faramarz Hojati}
\author[3]{Bahman Azarhoushang}
\author[2,*]{Ramin Aghababaei}

\affil[1]{Department of Computer Science, Aarhus University, Aarhus, Denmark; truongthitam@cs.au.dk (T.T.T.), panos@cs.au.dk (P.K.)}
\affil[2]{Department of Mechanical \& Production Engineering, Aarhus University, Aarhus, Denmark; jay@mpe.au.dk (J.A.), ra@mpe.au.dk (R.A.)}
\affil[3]{Institute of Precision Machining (KSF), Hochschule Furtwangen University, 78532 Tuttlingen, Germany; hofa@hs-furtwangen.de (F.H.); aza@hs-furtwangen.de (B.A.)}
\affil[*]{Corresponding author: ra@mpe.au.dk}

%% file: sections/00.abstract.tex
\begin{abstract}
The prediction of \emph{tool wear} helps minimize costs and enhance product quality in manufacturing. While existing data-driven models using machine learning and deep learning have contributed to the accurate prediction of tool wear, they often lack generality and require substantial training data for high accuracy. In this paper, we propose a new data-driven model that uses Bayesian Regularized Artificial Neural Networks (BRANNs) to precisely predict milling tool wear. BRANNs combine the strengths and leverage the benefits of artificial neural networks (ANNs) and Bayesian regularization, whereby ANNs learn complex patterns and Bayesian regularization handles uncertainty and prevents overfitting, resulting in a more generalized model. We treat both process parameters and monitoring sensor signals as BRANN input parameters. We conducted an extensive experimental study featuring four different experimental data sets, including the NASA Ames milling dataset, the 2010 PHM Data Challenge dataset, the NUAA Ideahouse tool wear dataset, and an in-house performed end-milling of the Ti6Al4V dataset. We inspect the impact of input features, training data size, hidden units, training algorithms, and transfer functions on the performance of the proposed BRANN model and demonstrate that it outperforms existing state-of-the-art models in terms of accuracy and reliability.
\end{abstract}

%% file: sections/01.introduction.tex
\section{Introduction}

Modern manufacturing requires efficient, flexible, and cost-effective processes to ensure a competitive advantage in the market and drive continuous improvement in production. As a \emph{milling tool} wears down, its performance and precision decline, causing a drop in product quality and a rise in production downtime and tool replacement costs. By accurately predicting tool wear, manufacturers can proactively plan tool replacements or maintenance, prevent tool failures and production interruptions, and thus maximize tools' lifespan and reduce operating expenses. Thus, the prediction of \emph{tool wear} is crucial in optimizing production processes to ensure cost-effectiveness and efficiency.

Various models have been developed and applied to predict tool wear. \emph{Physical models} consider various factors that contribute to tool wear, such as cutting forces, temperature distribution, material properties, tool geometry, cutting speed, and feed rate~\cite{airao2022analytical}, along with the underlying physics of the machining process, to simulate and predict the wear evolution of a cutting tool. In these models, every system component undergoes analysis to establish a physical failure model based on domain knowledge. Well-known physical models include the Taylor model~\cite{taylor1906art}, the Paris crack growth model \cite{colding1959wear}, and the Forman crack growth model~\cite{gilbert1950machining}. Recently, Ko and Koren~\cite{ko1989cutting} proposed a comprehensive physical model that relates clearance wear to the change of cutting force, making it an ideal tool for online tool wear assessment; Jawahir et al.~\cite{jawahir1995new} developed a new tool-life relationship that takes into account the effects of chip-groove parameters and tool coatings; Altintas et al.~\cite{altintas2008identification} introduced a cutting-force model that incorporates three dynamic cutting force coefficients associated with regenerative chip thickness, velocity, and acceleration terms; Kuttolamadom et al.~\cite{kuttolamadom2017high} utilized finite-element models (FEM) to assess the distributions of all physical variables during the cutting process; and Qin et al.~\cite{qin2009physics} presented a physics-based predictive model of cutting force in the Ultrasonic-vibration-assisted grinding (UVAG) of Ti. Still, albeit effective in predicting tool wear, these physical models have certain limitations: (1) they require expert knowledge to drive mathematical models, impeding their use by users lacking specialized expertise; (2) they often depend on predefined cutting parameters and tool geometries, delimiting their adaptability to new cutting conditions and tool designs; (3) they can be complex and computationally intensive, hence too time-consuming and resource-demanding for real-time or large-scale applications.

Recently, with the advent of the fourth industrial revolution or Industry~4.0, the manufacturing industries have undergone significant changes, incorporating advanced technologies such as the Internet of Things (IoT), artificial intelligence (AI), and robotics into the manufacturing process. This integration has resulted in a new era of production, where machines and systems are interconnected, leading to more efficient, flexible, and cost-effective manufacturing~\cite{tran2022machine}. A key sector that has embraced Industry~4.0 is tool condition monitoring, leading to smart monitoring systems~\cite{pimenov2023artificial} that utilize sensors and embedded systems to gather a large amount of data; such data, in turn, allow for real-time prediction of the tool condition using deep learning and machine learning.

Accordingly, significant attention has been paid to developing data-driven tool wear prediction models. Contrariwise to physical models, data-driven models rely on historical data collected during machining processes rather than on closed-form mathematical equations expressing physical laws. Therefore, these models capture complex relationships in the data, even when the underlying physics is not fully understood. Moreover, data-driven models can be effective with large datasets, which allow them to learn from a wide range of tool wear scenarios. For instance,  Ghosh et al.~\cite{ghosh2007estimation} developed a neural network-based sensor fusion model to estimate tool wear during the milling of C-60 steel, using the cutting force, spindle vibrations, and current signals for backward error-propagation learning. The estimation of tool wear using multiple signals exhibited good accuracy with respect to experimentally obtained values. Aghazadeh et al.~\cite{aghazadeh2018tool} used a Convolutional Neural Network (CNN) to predict tool wear by collecting force and current signals; this CNN reached an accuracy of~87.2\% for a system with spectral subtraction, Bayesian rigid regression, and support-vector regression. Huang et al.~\cite{huang2020tool} observed that the CNN was limited in its ability to learn from high-dimensional data, and hence adopted a Deep Convolutional Neural Network (DCNN) to estimate real-time tool wear for milling, acquiring force and vibration signals to extract features in multi-domains; tool wear estimated by DCNN showed a high accuracy with an RMSE of~0.0007998 mm. Qinglong et al.~\cite{an2020data} proposed a hybrid model, CNN-SBULSTM, to predict the remaining useful life (RUL) of milling tools; this model combines a CNN with a stacked bi-directional and uni-directional long short-term memory (SBULSTM) network; evaluated using datasets obtained from milling experiments, it could track tool wear progression and predict RUL with average prediction accuracy of up to~90\%. Likewise, Cai et al.~\cite{cai2020hybrid} presented a hybrid information model based on a long short-term memory network (LSTM) using both process information and sensor signal monitoring to predict tool wear; experimental results on the milling data set from NASA Ames and the data set reported in the 2010 PHM Data Challenge, showed the LSTM model to outperform several models such as linear regression (LR), support vector regression (SVR), multi-layer perceptron (MLP), and CNN, in prediction accuracy. To improve CNN accuracy, Xu et al.~\cite{xu2021deep} considered the significance of various features in the learning process, emphasizing the relevant ones while diminishing less useful ones during the tapping of AlSi7Mg by extracting forces and vibration signals from different channels and assigning respective weights to them; this method estimated tool wear with a maximum RMSE of~4.62. Liu et al.~\cite{liu2021meta} proposed a meta-invariant feature space (MIFS) learning model to predict tool wear under cross conditions and compared its performance to that of the DCNN and model agnostic meta-learning (MAML); showing the feasibility and accuracy of MIFS. Other studies on the application of data-driven models in manufacturing can be found in~\cite{sayyad2021data, mozaffar2022mechanistic}. Such data-driven models using machine learning and deep learning show their applicability in predicting tool wear; however, the models using \emph{conventional} machine learning usually suffer from over-fitting, leading to suboptimal accuracy and generalization, while those using \emph{deep learning} require a large amount of training data to obtain highly accurate and generalized models.

In this study, we propose a new data-driven model using Bayesian Regularized Artificial Neural Networks (BRANN) to overcome the limitations of the existing data-driven models for tool wear prediction. We use artificial neural networks (ANNs) to learn complex patterns from the data and Bayesian regularization to address uncertainty and prevent overfitting during the training process. By integrating these techniques, BRANN overcomes the shortcomings of conventional models, delivering enhanced performance in tool wear prediction. We validate the performance and applicability of BRANN through four different experimental data sets, namely the NASA Ames milling dataset, the 2010 PHM Data Challenge dataset, the NUAA Ideahouse tool wear dataset, and an in-house performed end-milling of the Ti6Al4V dataset. We investigate the influence of input features, training data size, hidden units, training algorithms, and transfer functions on the performance of BRANN and compare the results achieved by BRANN to those obtained from other existing models such as LR, SVR, MLP, CNN, LSTM, and MIFS, showcasing its superior accuracy and reliability.